\documentclass[submission, Phys]{SciPost}

\pdfoutput=1
\usepackage{cancel}
\usepackage{mathtools}
\usepackage[utf8]{inputenc} % input encodings (allow UTF-8 input)
\usepackage[T1]{fontenc} 	% selecting font encodings (use 8-bit T1 fonts)
\usepackage[english]{babel} % multilingual support (English language/hyphenation)

\usepackage[bitstream-charter]{mathdesign}
\usepackage{geometry} 		% flexible and complete interface to document dimensions
\usepackage{amsmath} 		% math package (American Mathematical Society)
\usepackage{mathtools} 		% math package (fixes various deficiencies of amsmath)
\usepackage{float} 			% floating objects such as figures and tables
\usepackage{graphicx} 		% enhanced support for graphics
\usepackage{tabularx} 		% tabulars with adjustable-width columns
\usepackage{booktabs} 		% professional-quality tables
\usepackage{color, xcolor} 	% foreground and background colour management
\usepackage{pdfpages} 		% inclusion of external multi-page PDF documents
\usepackage{extarrows} 		% extra arrows beyond those provided in amsmath
\usepackage{multirow} 		% create tabular cells spanning multiple rows
\usepackage{multicol} 		% intermix single and multiple columns
\usepackage{caption} 		% customising captions in floating environments
\usepackage{comment}
\usepackage{subcaption} 	% support for sub-captions (captions for subfigures)
\usepackage{enumitem} 		% control layout of itemize, enumerate, description
\usepackage{xspace} 		% define commands that appear not to eat spaces
\usepackage{stackrel} 		% enhancement to the \stackrel command
\usepackage{tikz} 			% create PostScript and PDF graphics
\usetikzlibrary{calc}
\usepackage{braket} 		% Dirac bra-ket notation
\usepackage{bm} 			% access bold symbols in maths mode
\usepackage{tensor} 		% typeset tensors (tensor-style super- and subscripts)
\usepackage{slashed} 		% slash through characters (Feynman slash notation)
\usepackage{siunitx} % SI units package (typesetting values with units)
\usepackage{lastpage} 		% reference last page
\usepackage{cite} 			% improved citation handling
\usepackage[normalem]{ulem} % package for underlining
\usepackage{fontawesome} 	% access to web-related icons
\usepackage{tocloft} 		% control over the typography of the TOC, LOF, and LOT
\usepackage{titlesec} 		% interface to sectioning commands (various title styles)
\usepackage{doi} 			% create correct hyperlinks for DOI numbers
\usepackage[most]{tcolorbox} 					% coloured and framed text boxes
\usepackage[capitalize]{cleveref} 	% intelligent cross-referencing
\usepackage[nottoc, notlot, notlof]{tocbibind} 	% add references/index/contents to TOC
\usepackage[ruled, vlined]{algorithm2e} 		% floating algorithm environment
\usepackage{makecell}
\usepackage{pifont}
\usepackage[dvipsnames]{xcolor}

\hypersetup{
	pdftitle={NAE, Statistically},
	pdfauthor={Ranit Das, Jonathan Ostertag-Henning, Tilman Plehn, and Lorenz Vogel},
    linktoc=section,
	breaklinks=true,			% splits links across lines
	colorlinks=true, 			% false: boxed links, true: colored links
	linkcolor={red!50!black}, 	% color of internal links (sections, pages, etc.)
	citecolor={blue!50!black}, 	% color of citation links (links to bibliography)
	urlcolor={blue!80!black} 	% color of URL links (external links)
} 
\DeclareSymbolFont{usualmathcal}{OMS}{cmsy}{m}{n}
\DeclareSymbolFontAlphabet{\mathcal}{usualmathcal}

\SetArgSty{textnormal}
\SetKwComment{Comment}{{\small\#}~}{}
\SetCommentSty{mycommfont}

\setitemize{itemsep=2pt, parsep=0pt} 				% adjust itemize environment
\setenumerate{itemsep=2pt, parsep=0pt} 				% adjust enumerate environment
\crefname{section}{Section}{Sections}
\crefname{equation}{Eq.}{Eqs.}
\crefname{figure}{Figure}{Figures}
\crefname{table}{Table}{Tables}
\Crefname{section}{Section}{Sections}
\Crefname{equation}{Equation}{Equations}
\Crefname{figure}{Figure}{Figures}
\Crefname{table}{Table}{Tables}
\definecolor{red_cb}{HTML}{e41a1c}
\definecolor{blue_cb}{HTML}{377eb8}
\definecolor{green_cb}{HTML}{4daf4a}
\definecolor{purple_cb}{HTML}{984ea3}
\definecolor{orange_cb}{HTML}{ff7f00}

\newcommand{\ie}{\text{i.e.}\;}

\newcommand{\clifford}{\text{Cl}_{1,3}(\R)}

\newcommand{\XLangle}{\Big\langle}
\newcommand{\XRangle}{\Big\rangle}

\newcommand{\qqquad}{\qquad\quad}
\newcommand{\qqqquad}{\qquad\qquad}

\newcommand\one{\leavevmode\hbox{\small1\normalsize\kern-.33em1}}
\newcommand{\abs}[1]{\lvert#1\rvert} 		% absolute value (single vertical lines)
\newcommand{\R}{\mathbb{R}} 				% real numbers
\newcommand{\cp}{\ensuremath{{\cal CP}}\xspace}

\newcommand{\invfb}{\,\, \mathrm{fb}^{-1}}

\newcommand{\ttbar}{\ensuremath{t\bar{t}}\xspace} 			% top-antitop pair
\newcommand{\arXiv}[2][]{%
	\ifthenelse{\equal{#1}{}}%
	{\href{http://arxiv.org/abs/#2}{arXiv:#2}}%
	{\href{http://arxiv.org/abs/#2}{arXiv:#2~[#1]}}}

\newcommand{\gev}{\text{GeV}}

\def\slashchar#1{\setbox0=\hbox{$#1$}           % set a box for #1
   \dimen0=\wd0                                 % and get its size
   \setbox1=\hbox{/} \dimen1=\wd1               % get size of /
   \ifdim\dimen0>\dimen1                        % #1 is bigger
      \rlap{\hbox to \dimen0{\hfil/\hfil}}      % so center / in box
      #1                                        % and print #1
   \else                                        % / is bigger
      \rlap{\hbox to \dimen1{\hfil$#1$\hfil}}   % so center #1
      /                                         % and print /
   \fi}

\newcommand{\tikznode}[2]{%
\ifmmode%
\tikz[remember picture,baseline=(#1.base),inner sep=0pt] \node (#1) {$#2$};%
\else
\tikz[remember picture,baseline=(#1.base),inner sep=0pt] \node (#1) {#2};%
\fi}

\def\mathswitchr#1{\relax\ifmmode{\mathrm{#1}}\else$\mathrm{#1}$\xspace\fi}
\def\mathswitch#1{\relax\ifmmode#1\else$#1$\xspace\fi}

\graphicspath{{./figures/}}

\begin{document}

\begin{center}
    {\Large\textbf{\color{scipostdeepblue}{Top Spin and CP from Equivariant Transformers}}}
\end{center}

\begin{center}\textbf{
Henning Bahl\textsuperscript{1},
Marco Menen\textsuperscript{2,3,4},
Antoine Petitjean\textsuperscript{1}, and
Tilman Plehn\textsuperscript{1,5}
}\end{center}

\begin{center}
{\bf 1} Institut für Theoretische Physik, Universit\"at Heidelberg, Germany \\
{\bf 2} Institut für Theoretische Physik, Leibniz Universität Hannover, Germany \\
{\bf 3} Physikalisch-Technische Bundesanstalt, Braunschweig, Germany \\
{\bf 4} Department of Physics, TU Dortmund University, Germany \\
{\bf 5} Interdisciplinary Center for Scientific Computing (IWR), Universität Heidelberg, Germany
\end{center}

\begin{center}
    \today
\end{center}
 
\section*{\color{scipostdeepblue}{Abstract}}
\textbf{\boldmath{Measurements of spin correlations and CP are an exciting and challenging part of the precision-LHC program. We first show how L-GATr-slim, as a Lorentz-equivariant transformer, significantly boosts the spin analyzing power for hadronic top quarks. We then show how adding a pseudoscalar channel to its architecture allows us to learn and interpret optimal observables and their CP properties. We demonstrate this for WBF-Higgs production at parton level and top pair production including detector effects.}}

\vspace{10pt}
\noindent\rule{\textwidth}{1pt}
\tableofcontents\thispagestyle{fancy}
\noindent\rule{\textwidth}{1pt}
\vspace{10pt}

\clearpage
%%%%%%%%%%%%%%%%%%%%%%%%%%%%%%%%%%%%%%%%%%%%%%%%%%
\section{Introduction}
\label{sec:introduction}

Top physics is at the heart of the LHC-precision program, as the top, with its large mass and large production rate, provides a unique experimental probe of fundamental particle properties. Its short lifetime implies that it decays before hadronic bound states can form, so the decay transfers information about the top polarization directly to its decay products~\cite{Bigi:1986jk}. Top-pair spin correlations were first measured in purely leptonic decays. Only recently have they been extended to lepton-hadronic decays~\cite{CMS:2013roq,ATLAS:2014abv,ATLAS:2015igw,CMS:2016piu,ATLAS:2019zrq,CMS:2019nrx,ATLAS:2014aus,CMS:2024zkc,CMS:2025brx}. Independent of a potential interpretation in a quantum mechanics framework~\cite{Abel:1992kz,Abel:2025wwa,Bechtle:2025ugc}, spin correlations in top pair production conclusively probe \cp violation in the top sector~\cite{Lamba:2026jfm,Lamba:2026sni,Erdogan:2026cbe,Perelstein:2026dbu}. 

The cleanest spin analyzers in the top decay are the charged lepton and the down-type quark~\cite{Jezabek:1994zv,Brandenburg:2002xr}. For hadronic decays, the bottleneck is the identification of the down quark. A naive first approach simply identifies the softer light-quark jet with the down quark. Better results can be achieved with a calibrated down-quark probability~\cite{Atwood:2000tu,Tweedie:2014yda,Dong:2024xsb,Dong:2024xsg}, but remain far from optimal. 

Modern machine learning is transforming every aspect of LHC physics~\cite{Plehn:2022ftl,Krause:2026ayh}, where the state-of-the-art architecture is transformers~\cite{Mikuni:2021pou,Qu:2022mxj,Qiu:2022xvr,DiBello:2022iwf,Finke:2023veq,Butter:2023fov,Leigh:2024ked,Brehmer:2024yqw,Butter:2024zbd,Favaro:2025pgz,Petitjean:2025zjf,Breso-Pla:2026tlz,Favaro:2026amj,Vigl:2026bah}. LHC physics is governed by fundamental symmetries, which can be incorporated explicitly to make transformers more efficient and accurate~\cite{Breso-Pla:2026tlz}. Lorentz-equivariant transformers~\cite{Favaro:2026amj} can be constructed using geometric algebra (L-GATr)~\cite{Brehmer:2024yqw,Petitjean:2025zjf} or boosting between separate reference frames (LLoCa)~\cite{Favaro:2025pgz}. We will improve the polarization measurement in hadronic top decays using a state-of-the-art L-GATr-slim classifier~\cite{Petitjean:2025zjf}.

\cp violation is introduced into top pair production via dimension-6 SMEFT operators~\cite{Grzadkowski:2010es,Brivio:2017vri}. Specifically, a complex Wilson coefficient of the chromomagnetic dipole operator generates \cp-odd interference. It can be tracked via \cp-odd observables, requiring spin information and built via the Levi-Civita tensor~\cite{Antipin:2008zx,Gupta:2009wu,Han:2009ra,Hayreter:2015ryk,Bernreuther:2017yhg,Brehmer:2017lrt}. Optimal observables building on this \cp-odd observable exploit the full phase space correlations. At the detector, these optimal observables are hard to construct explicitly~\cite{Brehmer:2017lrt},
motivating the use of ML-methods in the search for \cp violation~\cite{Bortolato:2020zcg,Bahl:2021dnc,Barman:2021yfh,Bhardwaj:2021ujv,Butter:2021rvz,Hall:2022bme,Ackerschott:2023nax,Bahl:2023qwk,Barrue:2023ysk,Cruz:2024grk,Esmail:2024gdc,Bahl:2025jtk,Boto:2026gzj}. We will again use a Neyman-Pearson-optimal L-GATr-slim classifier, but now to distinguish \cp-conjugate WBF-Higgs and top pair events and construct the log-likelihood ratio~\cite{Cranmer:2015bka} as an optimal \cp observable.

In Sec.~\ref{sec:top_quark}, we first review the extraction of the top polarization in hadronic decays and the construction of \cp-odd observables for top pairs. In Sec.~\ref{sec:downtagging} we use L-GATr-slim~\cite{Petitjean:2025zjf} to tag down-type quarks in hadronic top decays, which boosts the spin analyzing power significantly. In Sec.~\ref{sec:tt_smeft} we modify L-GATr-slim to separate scalar and pseudoscalar outputs and demonstrate how this leads to learned \cp-even and \cp-odd observables. We validate our result for WBF-Higgs production at parton level and for top pair production including detector effects. We conclude in Sec.~\ref{sec:conclusion}.

%\clearpage
%%%%%%%%%%%%%%%%%%%%%%%%%%%%%%%%%%%%%%%%%%%%%%%%%%
\section{Top polarization and CP-violation}
\label{sec:top_quark}

In top decays, the subsequent $W$-decay determines the final state,
\begin{align}
    t\to W^+ b \to 
    \begin{cases}
        \ell^+ \nu b & \text{(leptonic)} \\
        \bar d u b & \text{(hadronic)} \; .
    \end{cases}
\end{align}
The degree of top spin polarization $P$ is defined through the helicity angle $\theta_i$ between the decay particle $i$ and the top spin axis,
\begin{align}
\label{eq:helicity_angle}
    \frac{1}{\Gamma} \frac{d\Gamma}{d\cos\theta_i} = \frac{1 + \beta_i P \cos\theta_i}{2} \;.
\end{align}
This formula also defines the spin analyzing power $\beta_i$ as the amount of polarization information carried by the decay product $i$. To leading order it is~\cite{Tweedie:2014yda}
\begin{align}
\label{eq:spin_analyzing_power}
    \beta_i= 
    \begin{cases} 
            1\; & i = \ell^+,\bar d \\
        -0.34\; & i = \nu, u \\
        -0.40\; & i = b \; .
    \end{cases} 
\end{align}
For the $\bar{t}$ decay, the sign of $\beta_i$ is flipped. Tagging a down quark is challenging, and the associated uncertainties can wash out the polarization information. This points to the easily identified and reconstructed lepton, at the price of significantly lower statistics, especially when correlating decays of top pairs. 

%%%%%%%%%%%%%%%%%%%%%%%%%%%%%%%%%%%%%%%%%%%%%%%%%%
\subsubsection*{Hadronic top decays}

We estimate the spin analyzing power from the unit 3-momentum of the down quark $\hat k_d$. For hadronic decays with a $b$-tag, the down quark assignment is
\begin{align}
 d \in \left\{ q_\text{soft}, q_\text{hard} \right\} \; ,
\end{align}
ordered by energy in the top rest frame.  Following Eq.\eqref{eq:helicity_angle}, the top decay products are distributed jointly over phase space in the top rest frame with a spin axis $\hat{n}$ and this down quark assignment,
\begin{align}
\label{eq:polarized_distribution}
    f_P(x, d) 
    &= f_0(x, d) \left( 1 + P~\hat{k}_d \cdot \hat{n} \right) \notag \\
    &\equiv f_0(x, d) \left( 1 + P\cos \theta_d \right) \; .
\end{align}
This corresponds to the optimal $\beta_d = 1$. The marginal kinematic distribution is given by the superposition
\begin{align}
\label{eq:polarized_distribution_split}
    f_P(x) 
    &= f_P(x, d = q_\text{soft}) +  f_P(x, d = q_\text{hard}) \notag \\
    &= f_0(x,d = q_\text{soft}) \left( 1 + P~\hat{k}_\text{soft} \cdot \hat{n} \right) + f_0(x,d = q_\text{hard}) \left( 1 + P~\hat{k}_\text{hard} \cdot \hat{n} \right) \notag \\
    &= f_0(x)\Big[ 1 + P\left(p(d = q_\text{soft}|x)\, \hat{k}_\text{soft} + p(d = q_\text{hard}|x)\, \hat{k}_\text{hard}\right) \cdot \hat n \Big] \notag \\
    \text{with} &\qquad 
    p(d = q_\text{soft/hard}|x) = \frac{f_0(x,d = q_\text{soft/hard})}{f_0(x)} \; .
\end{align}
We define the optimal down-probe vector to recover the form of Eq.\eqref{eq:polarized_distribution},
\begin{align}
\label{eq:optimal_spin_analyzer}
    \vec k_\text{opt}(x) &= p(d = q_\text{soft} | x)\, \hat{k}_\text{soft} + p(d = q_\text{hard} | x)\, \hat{k}_\text{hard} \notag \\
    \Rightarrow \qquad f_P(x) &= f_0(x) \; \Big[ 1 + P \; \vec k_\text{opt}(x) \cdot \hat{n} \Big] \; .
\end{align}
We approximate this optimal vector by a unit probe for the down-type 3-momentum $\tilde k_d(x)$ and the corresponding helicity angle $\tilde \theta_d$. Its kinematic distribution becomes 
\begin{align}
\label{eq:probe_distribution}
    \frac{1}{\Gamma}\frac{d\Gamma}{d \cos \tilde \theta_d} 
    &= \int dx f_0(x)~\Big[ 1 + P~ \vec k_\text{opt}(x) \cdot \hat n \Big]~ \delta(\cos \tilde \theta_d - \tilde k_d(x) \cdot \hat n) \notag \\
    &= \int dx f_0(x)~\Big[ 1 + P~\Big(\tilde k_d(x) \cdot \vec k_\text{opt}(x)\Big)\Big(\tilde k_d(x) \cdot \hat n\Big)\Big]~ \delta(\cos \tilde \theta_d - \tilde k_d(x) \cdot \hat n) \notag \\
    &= \frac{1 + P \; \XLangle \tilde k_d(x) \cdot \vec k_\text{opt}(x) \XRangle_x  \cos \tilde \theta_d}{2} \; .
\end{align}
Here we use the isotropy of the distribution to integrate out the part of $\vec k_\text{opt}$ orthogonal to $\tilde k_d$. The corresponding spin analyzing power is 
\begin{align}
\label{eq:beta_probe}
    \tilde \beta_d 
    = \XLangle \tilde k_d(x) \cdot \vec k_\text{opt}(x) \XRangle_x 
    \leq \XLangle \abs{\vec k_\text{opt}(x)} \XRangle_x \equiv \beta_\text{opt} \; .
\end{align}
A simple historical approximation is to identify the down quark with the softer $W$-decay quark
\begin{align}
\label{eq:down_pt}
    \tilde k_d = \hat k_\text{soft}
    \qquad \Rightarrow \qquad \beta_\text{soft} \approx 0.5 \; .
\end{align}
Better probes can be constructed by estimating $p(d = q_\text{soft/hard}|x)$. Using the angular distribution of light quarks in the $W$ rest frame with respect to the $b$ momentum gives~\cite{Tweedie:2014yda}
\begin{align}
    \beta_{c_{W\text{hel}}} \approx 0.64  \; .
\end{align}
Modern ML allows us to estimate the probability from the constituent kinematics as well as their charge and particle IDs. A graph neural network based on ParticleNet~\cite{Qu:2019gqs} yields~\cite{Dong:2024xsg}
\begin{align}
    \beta_\text{GNN} \approx 0.68 \; .
\end{align}
In Sec.~\ref{sec:downtagging} we will look at ways to improve this down-quark probe using state-of-the art transformers.

%%%%%%%%%%%%%%%%%%%%%%%%%%%%%%%%%%%%%%%%%%%%%%%%%%
\subsubsection*{Top pairs and \cp violation}

While individual tops at the LHC are unpolarized, the spins in (associated) top pair production are correlated,
\begin{align}
\label{eq:top_spin_correlation}
    \frac{1}{\sigma} \frac{d^2\sigma} {d\cos\theta_1\, d\cos\theta_2} 
    = \frac{1 + B_1 \cos\theta_1 + B_2 \cos\theta_2 - C \cos\theta_1 \cos\theta_2}{4} \; .
\end{align}
$B_1$ and $B_2$ are polarization vectors and $C$ is the spin correlation matrix. Experimental measurements mostly use the lepton-lepton decays~\cite{CMS:2013roq,ATLAS:2014abv,ATLAS:2015igw,CMS:2016piu,ATLAS:2019zrq,CMS:2019nrx}. The first study of the lepton-hadron channel performed worse~\cite{ATLAS:2014aus}, but new measurements benefit from more data and ML~\cite{CMS:2024zkc,CMS:2025brx}. 

\cp violation can be introduced into top pair production via a dimension-6 operator with a complex coefficient~\cite{Brivio:2019ius},
\begin{align}
    \mathcal{O}_{tG} = \left( \overline{Q} \sigma^{\mu \nu} T^A t \right) \tilde{\phi} \, G_{\mu \nu}^A   
    \qquad \text{with} \qquad 
    \text{Im}(c_{tG}) \neq 0 \; .
\end{align}
This interaction leads to a \cp-violating interference in the \ttbar cross section
\begin{align}
\label{eq:matrix-element-cpodd}
    |\mathcal{M}|^2 = |\mathcal{M}_\text{SM}|^2 + 2 c_{tG} \text{Re}(\mathcal{M}_\text{SM}^* \mathcal{M}_\text{CP}) + c_{tG}^2 |\mathcal{M}_\text{CP}|^2 \; .
\end{align}
It can be probed by \cp-odd observables of the general form~\cite{Han:2009ra,Brehmer:2017lrt}
\begin{align}
\label{eq:epsilon-cpodd}
    O \propto \epsilon_{\mu\nu\rho\sigma} p_1^\mu p_2^\nu p_3^\rho p_4^\sigma \; .
\end{align}
An additional \cp-even prefactor can be adapted to the kinematics induced by the \cp-violating operator. This observable requires four independent 4-momenta, so for a $2 \to 2$ process we need spin information, for instance, decay 4-momenta. At the parton level
\begin{align}
\label{eq:epsilon-semilep}
    p_1 = k_t
    \qqqquad p_2 = k_{\bar{t}}
    \qqqquad p_3 = k_\ell 
    \qqqquad p_4 = k_{\ell/d} \; ,
\end{align}
where $p_4$ depends on the decay channel. In the \ttbar rest frame, $k_t = (\sqrt{s}/2, \pm \vec{k}_t)$ gives the triple product
\begin{align}
    O \propto - \frac{1}{2} \sqrt{s}~\epsilon_{ijk} k_t^i k_\ell^j k_{\ell/d}^k = - \frac{1}{2} \sqrt{s}~\vec{k}_t \cdot (\vec{k}_\ell \times \vec{k}_{\ell/d}) \; .
\end{align}
It is directly related to the antisymmetric part of the spin correlation matrix $C$ in Eq.\eqref{eq:top_spin_correlation}~\cite{Bernreuther:2017yhg}. 

%%%%%%%%%%%%%%%%%%%%%%%%%%%%%%%%%%%%%%%%%%%%%%%%%%
\section{Hadronic top-decay polarization}
\label{sec:downtagging}

To illustrate our reconstruction of the top polarization by classifying the two $W$-decay jets, we generate top pair events at the parton level using MadGraph~\cite{Alwall:2014hca} with fixed polarizations
\begin{align}
pp \to (t\bar{t})_{P,P'} 
\qquad \text{with} \qquad 
PP' \in \left\{ 00, LL, RR, LR, RL \right\} \; ,
\end{align}
where $0$ stands for unpolarized. We simulate lepton-hadron decays with MadSpin, evenly split into hadronic top and anti-top events. We shower the events with Pythia8~\cite{Bierlich:2022pfr} and simulate detector effects with Delphes3~\cite{deFavereau:2013fsa}. We reconstruct jets with FastJet~\cite{Cacciari:2011ma} using the C/A algorithm with $\Delta R = 0.4$ and $1.5$.

We require each event to have at least one fat jet, exactly one hard lepton, and exactly two b-tagged jets. The $p_T$-leading fat jet is required to have
\begin{align}
 p_T > 250~\gev
 \qquad \text{and} \qquad 
 |\eta| < 3 \; .
\end{align}
We de-cluster the fat jet~\cite{Plehn:2009rk,Plehn:2010st} again using FastJet, following Ref.~\cite{Dong:2024xsg}. We revert the clustering sequence until we reach at least three subjets with $m_\text{subjet} < 30~\text{GeV}$. We identify the $b$-subjet and the two light subjets by matching them to the top decay partons within $\Delta R < 0.4$. We modify the mass window of Ref.~\cite{Dong:2024xsg} to~\cite{CMS:2009lxa}
\begin{align}
 m_\text{fat} \in [120, 220]~\gev \; .
\end{align}
We then reconstruct the leptonic top decay using the $m_W$ criterion to extract $p_{z,\nu}$ and combine the reconstructed $W$ with the second $b$-tagged jet in the event, using the same top mass window.

After reconstruction, we are left with $1.98 \cdot 10^6$ unpolarized top jets. We use $80\%$, $10\%$, and $10\%$ of them for training, validation, and testing. We use $2.41 \cdot 10^5$ polarized top jets for evaluation, with an equal share of LL, LR, RL, and RR polarizations.

We store the set of 4-momenta of a single top in fixed order: 
\begin{enumerate}
 \item the hadronic top candidate; 
 \item the $b$-tagged subjet;
 \item the light subjets, ordered by their energy in the hadronic top rest frame; and
 \item all constituents grouped by the subjet in which they appear, descending in $p_T$.
\end{enumerate}
We ignore all information about the leptonically decaying top in this section. We then add a set of precomputed features per particle that are standard for jet tagging,
\begin{align}
\label{eq:scalar-inputs}
\Big\{ \;
\log p_T, \log E, \Delta \log p_T, \Delta \log E, \Delta \phi, \Delta \eta, \Delta R \; \Big\} \; .
\end{align}
Finally, we encode the jet structure, particle ID, and charge. The 4-momenta correspond to vector inputs and the remaining inputs are scalars. We refer to App.~\ref{app:inputs-hyperparameters} for details.

%%%%%%%%%%%%%%%%%%%%%%%%%%%%%%%%%%%%%%%%%%%%%%%%%%
\subsubsection*{L-GATr-slim}

We employ the Lorentz-equivariant transformer implementation L-GATr-slim~\cite{Petitjean:2025zjf} to improve the hadronic top polarimetry. For applications under realistic conditions, it is important to train equivariant transformers including a slight symmetry breaking, that we explain in App.~\ref{app:symmetry-group}.

Typically, a transformer input token contains all information about one (reconstructed) particle. L-GATr embeds all inputs in the Clifford algebra $\clifford$, as the Lorentz group has a natural representation that commutes with its operators. The 16-dimensional multivector elements of the algebra include a scalar, vector, bivector, axial vector, and pseudoscalar,
\begin{align}
\label{eq:lgatr-algebra}
(s,v,b,a,p) \in \R^1 \times \R^4 \times \R^6 \times \R^4 \times \R^1 \; .
\end{align}    
An input particle is described by its 4-momentum $v$ and a set of scalar features. The outputs of the comprehensive L-GATr implementation are the elements of the geometric algebra. LHC tasks that only require scalars and vectors motivate the L-GATr-slim implementation: it drops all other multivector elements without any performance loss across regression, classification, and generation~\cite{Petitjean:2025zjf}.

%%%%%%%%%%%%%%%%%%%%%%%%%%%%%%%%%%%%%%%%%%%%%%%%%%
\subsubsection*{Down-type tagger}

Following Sec.~\ref{sec:top_quark}, an optimal extraction of the top polarization requires tagging the down-type quark, $d = q_\text{soft/hard}$. We train the L-GATr-slim classifier on $1.58\cdot 10^6$ unpolarized jets and give all hyperparameters in App.~\ref{app:inputs-hyperparameters}. We average the results over five runs and obtain the uncertainty
\begin{align}
\label{eq:network-uncertainty}
    \sigma = \sqrt{\sigma_\text{seed}^2 + \sigma_\text{re-sample}^2}
\end{align}
from the five independent seeds and 50 bootstrap re-samplings of the test dataset.

%----------------------------------------
\begin{figure}[b!]
    \includegraphics[width=0.495\textwidth]{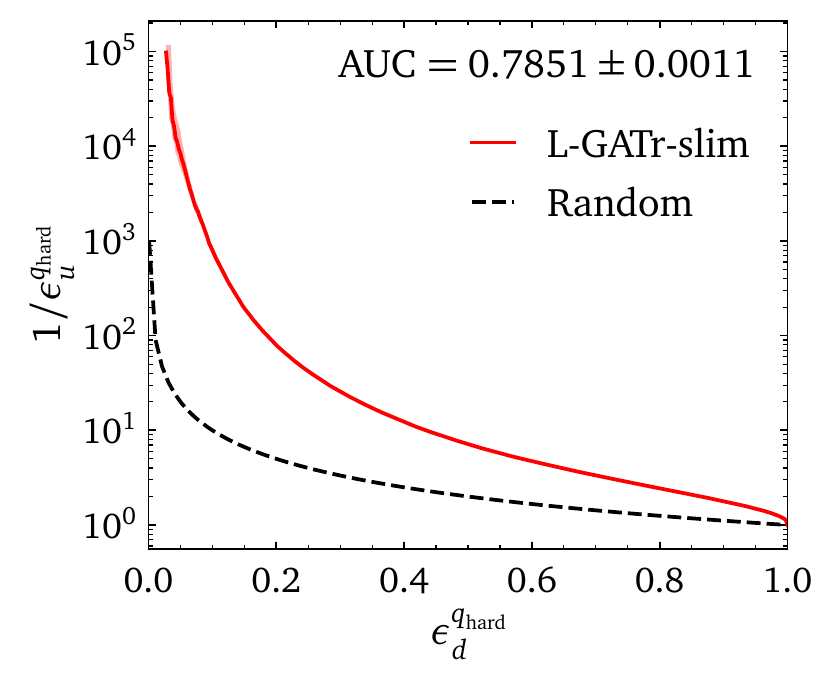}
    \includegraphics[width=0.495\textwidth]{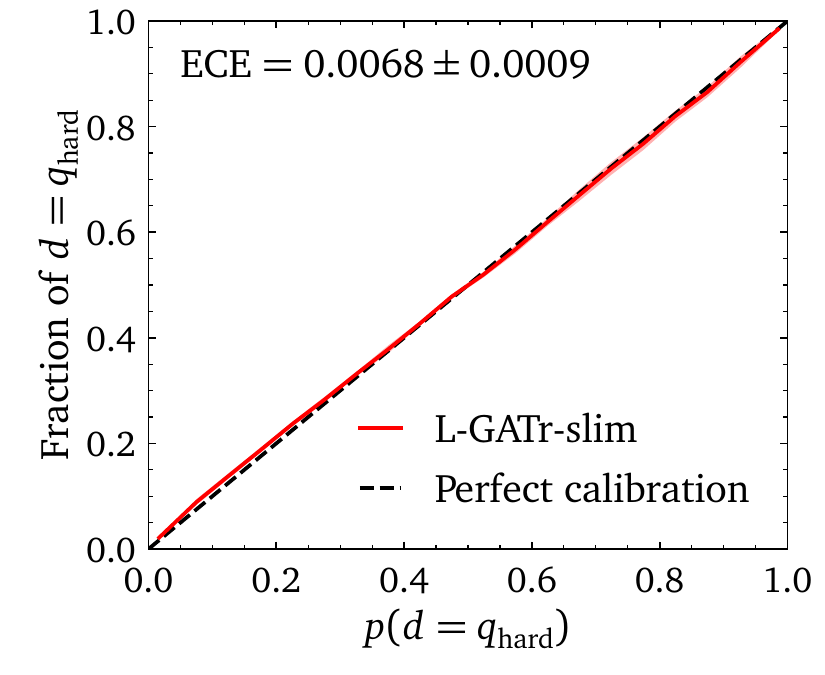}
    \caption{Left: Background rejection rate of the down tagger over the signal efficiency. Right: calibration of the L-GATr-slim tagger, \ie fraction of test events with $d = q_\text{hard}$ in bins of the predicted probability $p(d = q_\text{hard})$.}
    \label{fig:downtag-roc}
\end{figure}
%----------------------------------------

In the left panel of Fig.~\ref{fig:downtag-roc} we see that an identification of the flavor of the $W$-decay jets is possible, with an AUC of $0.785 \pm 0.001$. We evaluate the tagger output as the down-type probability of the harder $W$-decay jet. The optimal probe of Eq.\eqref{eq:optimal_spin_analyzer} requires this output to be a calibrated probability. 

In the right panel of Fig.~\ref{fig:downtag-roc} we bin the test events in the predicted $p(d = q_\text{hard})$ in 20 bins and show the fraction of events in each bin for which the down quark is indeed the harder $W$-decay quark. The curve follows the diagonal over the full range, which means the classifier is correctly calibrated.

Next, we use the classifier output to build the down momentum probe $\tilde{k}_d$ and determine the spin analyzing power of our new approach. We define the L-GATr-slim probe as 
\begin{align}
\label{eq:lgatr-probe}
    \tilde k_d  = \frac{p_\text{L-GATr}(d=q_\text{hard})~\hat{k}_\text{hard} + p_\text{L-GATr}(d=q_\text{soft})~\hat{k}_\text{soft}}{\abs{p_\text{L-GATr}(d=q_\text{hard})~\hat{k}_\text{hard} + p_\text{L-GATr}(d=q_\text{soft})~\hat{k}_\text{soft}}} 
\end{align}
and reach the currently leading spin-analyzing power 
\begin{align}
    \beta_\text{L-GATr}
    = \frac{1}{N_{\text{test}}} \sum_{\text{test}} \tilde k_d \cdot \hat k_d 
    = 0.721 \pm 0.001 \; .
\end{align}

We can use the classifier-based down quark probe to predict the helicity angle and compare it with the parton-level truth. The helicity angle is only meaningful for polarized tops. We pass the left-handed and right-handed polarized tops to the trained classifier and calculate again $\tilde k_d$ according to Eq.\eqref{eq:lgatr-probe}. The helicity angle is then
\begin{align}
    \cos \tilde \theta_d = \tilde k_d \cdot \hat{n} \; ,
\end{align}
where we choose the top spin axis $\hat{n}$ as the 3-momentum of the leptonic top in the hadronic top rest frame, although different conventions exist. 

%----------------------------------------
\begin{figure}[t]
    \includegraphics[width=0.495\textwidth]{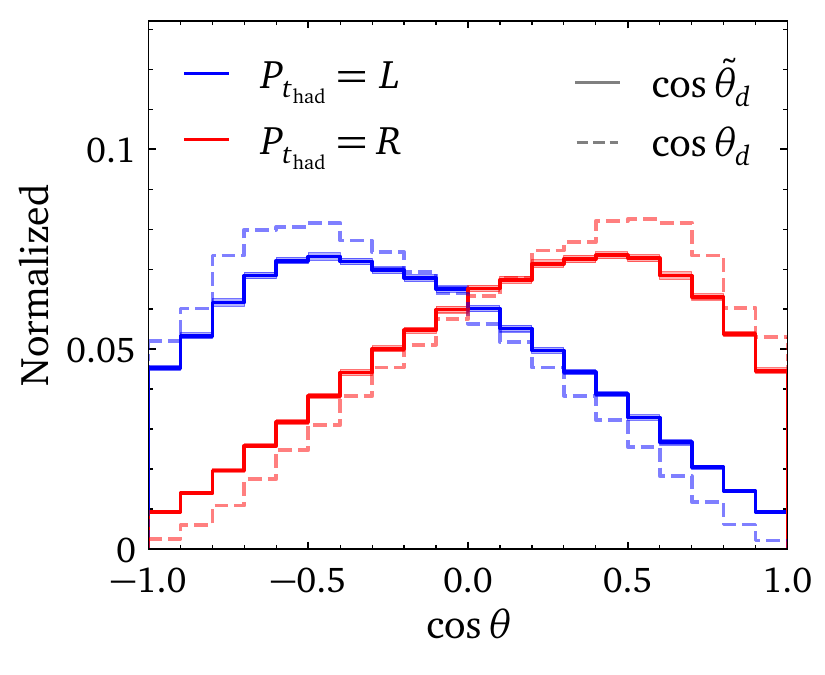}
    \includegraphics[width=0.495\textwidth]{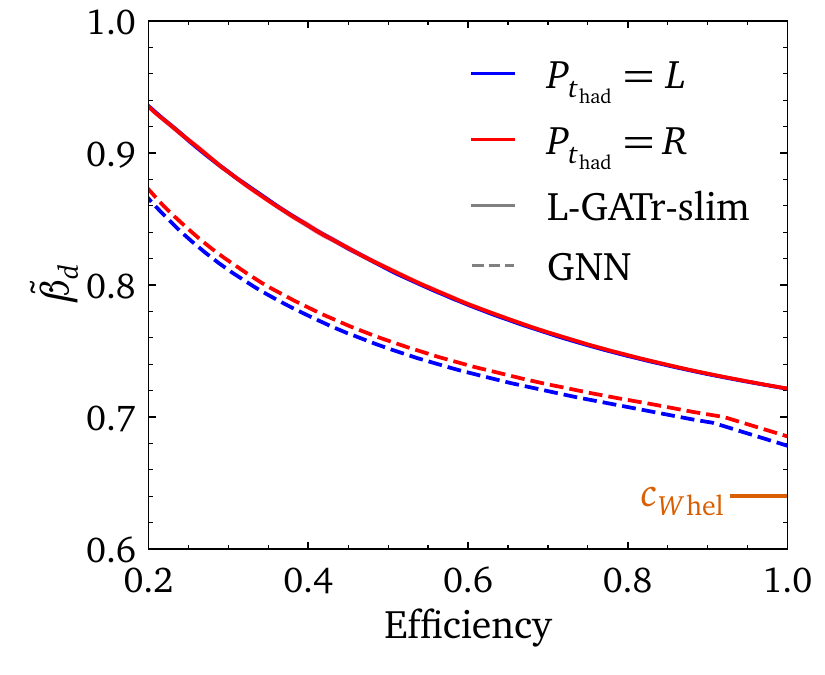}
    \caption{Left: helicity angle of the down-type direction predicted from L-GATr-slim, compared to the true reconstructed down-type direction. Right: hadronic spin analyzing power as a function of the network efficiency for L-GATr-slim, compared to the GNN used in Ref.~\cite{Dong:2024xsg} and the $W$-helicity probe of Ref.~\cite{Tweedie:2014yda} at full efficiency.}
    \label{fig:lgatr_helicity}
\end{figure}
%----------------------------------------

In the left panel of Fig.~\ref{fig:lgatr_helicity} we show the distribution of the helicity angle, split into hadronic top polarizations. While the parton-level expectation is a linear trend with a positive (negative) gradient for right-(left-)handed tops, the distributions of $\cos \theta_d$ drop at the edges at the detector level. The distributions using the L-GATr-slim tagger reproduce the peaks around $\pm 0.5$, the reduced slope of the linear trend reflecting the imperfect down-tagging.

Finally, we follow Ref.~\cite{Dong:2024xsg} and report the spin analyzing powers at different signal efficiencies. The efficiency is defined as the fraction of events sorted by confidence, \ie the events where $\abs{p_\text{L-GATr}(d = q_{\text{hard}}) - 0.5}$ is the largest, such that a smaller efficiency leads to larger $\beta_\text{L-GATr}$. In the right panel of Fig.~\ref{fig:lgatr_helicity} we show the spin analyzing powers, individually extracted for left-handed and right-handed tops, as a continuous function of the efficiency. The L-GATr-slim tagger does not develop a preference for one of the polarization states and tags down-type quarks from left-handed and right-handed tops with the same accuracy at each efficiency. The top $20\%$ events with the highest network confidence achieve an average spin analyzing power of $0.935 \pm 0.002$. 

\clearpage
%%%%%%%%%%%%%%%%%%%%%%%%%%%%%%%%%%%%%%%%%%%%%%%%%%
\section{\cp observable from equivariance}
\label{sec:tt_smeft}

Moving from the polarization of a single top to the spin correlation of top pairs as a probe of \cp violation, we use the same L-GATr-slim classifier to construct an optimal \cp-odd observable. In contrast to the original L-GATr, the default version of L-GATr-slim only includes scalars, invariant under $\text{O}(1,3)$, and vectors, covariant under $\text{O}(1,3)$. To extract \cp information, we need to process  \cp-even and \cp-odd information separately.

%%%%%%%%%%%%%%%%%%%%%%%%%%%%%%%%%%%%%%%%%%%%%%%%%%
\subsection{\cp in L-GATr-slim}
\label{subsec:lgatr-pseudo}

Similarly to the original L-GATr, we now include the pseudoscalar multivector element from Eq.\eqref{eq:lgatr-algebra} in L-GATr-slim. We refer to Ref.~\cite{Petitjean:2025zjf} for a description of the L-GATr-slim layers and to App.~\ref{app:symmetry-group} for more details on the symmetry group and just summarize the extended structure:
\begin{itemize}
\item In the linear layer, pseudoscalars do not include a bias which would be \cp-invariant,
\begin{align}
    \text{Linear} (x) 
    = 
    \text{Linear} 
    \begin{pmatrix} s\\v\\p \end{pmatrix} 
    = \begin{pmatrix} \text{Linear}_s(s)\\\text{Linear}_v(v)\\\text{Linear}_p(p) \end{pmatrix} 
    = \begin{pmatrix} w_s\cdot s + b_s  \\ w_v\cdot v \\  w_p \cdot p \end{pmatrix} \; .
\end{align}
\item The nonlinearity is a modified version of the Gated Linear Unit (GLU)~\cite{dauphin2017language} used in L-GATr-slim. We apply non-linear activation functions to Lorentz invariants,
\begin{align}
    \text{Act}\begin{pmatrix}s\\v\\p\end{pmatrix} = \begin{pmatrix}
        \text{GELU}\left(\text{Linear}_{s}(s)\right) \\
        \text{Sigmoid}\left(\langle \text{Linear}_{v,1}(v),\text{Linear}_{v,2}(v)\rangle\right) \\
        \text{GELU}\left((\text{Linear}_{p,1}(p))^2+b_p\right)
    \end{pmatrix} \; ,
\end{align}
and multiply them element-wise with a linear layer
\begin{align}
    \text{GLU}\begin{pmatrix}s\\v\\p\end{pmatrix} = \text{Act}\begin{pmatrix}s\\v\\p\end{pmatrix}
    \odot \text{Linear}\begin{pmatrix}s\\v\\p\end{pmatrix}
    \; .
\end{align}
\item The layer norm is modified over the original L-GATr-slim: each multivector channel is normalized separately for numerical stability instead of using a shared norm,
\begin{align}
    \text{RMSNorm}(x) = \frac{x}{\sqrt{\dfrac{\sum_c x_c^2}{c_x} + \epsilon}} 
    \qquad \text{with} \qquad 
    x = s,v,p \quad \text{and} \quad v^2 = \left| \langle v,v\rangle \right| \; ,
\end{align}
where $c_x$ is the number of $x$ channels.
\item The attention uses the fact that the Minkowski inner product and the product of two pseudoscalars are both Lorentz-invariant,
\begin{align}
    \text{Attention}\begin{pmatrix}q_s,k_s,v_s\\q_v,k_v,v_v\\q_p,k_p,v_p\end{pmatrix}_i 
    &= \sum_{j=1}^{n_t} \text{Softmax}_j\left(A_{ij}\right) \begin{pmatrix}v_{s}\\v_{v}\\v_p\end{pmatrix}_j \notag \\
    \text{with} \qquad 
    A_{ij} 
    &= \frac{\sum_c q_{s,ic}k_{s,jc} 
    + \sum_c \langle q_{v,ic},k_{v,jc}\rangle
    +\sum_c q_{p,ic}k_{p,jc}}{\sqrt{n_s + 4n_v +n_p}} \; .
\end{align}
\item To allow the network to separate pseudoscalar information from the vectors, we build \cp-odd observables following Eq.\eqref{eq:epsilon-cpodd}
\begin{align}
    \det[v_1,v_2,v_3,v_4] = -\epsilon_{\mu\nu\sigma\rho}v_1^\mu v_2^\nu v_3^\sigma v_4^\rho \; ,
\end{align}
We change the number of vectors from $c_v$ to four times the number of pseudoscalars with a learned projection $w$, and we compute a 4-determinant for each pseudoscalar
\begin{align}
    \text{PseudoProjection} \begin{pmatrix}s\\v\\p\end{pmatrix} = \begin{pmatrix}s\\v\\p + \text{det}[w\cdot v] \end{pmatrix}
\end{align}
Additional operations mixing the three multivector elements are not needed, thanks to the expressivity of the attention.
\end{itemize}
We stack the layers like in L-GATr-slim, with the PseudoProjection acting on the output of the attention,
\begin{align}
    \bar x &= \text{RMSNorm}(x) \notag\\
    \text{AttentionBlock}(x) &= \text{Attention}\left(\text{Linear}(\bar x),\text{Linear}(\bar x), \text{Linear}(\bar x)\right)\notag\\
    \text{MLPBlock}(x) &= \text{Linear}\circ \text{GLU}(\bar x) + x\notag\\
    \text{Block}(x) &= \text{MLPBlock}(\text{PseudoProjection}\circ  \text{AttentionBlock}(x)+x)\notag\\
    \text{L-GATr-slim}(x) &= \text{Linear}\circ\text{Block}\circ\text{Block}\circ\cdots\circ\text{Block}\circ\text{Linear}(x) \; .
\end{align}
This architecture splits the input variables into scalars and pseudoscalars such that the outputs $s(x)$ and $p(x)$ are exactly \cp-even and \cp-odd. We define the final output as the sum of the two because any real function can be written as the sum of even and odd contributions.

%%%%%%%%%%%%%%%%%%%%%%%%%%%%%%%%%%%%%%%%%%%%%%%%%%
\subsubsection*{Learned \cp observable}

Instead of constructing \cp-odd observables by hand, we can use a classifier. Consider two \cp-linked probability distributions related by
\begin{align}
\label{eq:cp-link}
    f_\pm(x) 
    \; \overset{\cp}{\longleftrightarrow} \; 
    f_\mp(x) \; .
\end{align}
An optimal classifier between these two phase space densities satisfies
\begin{align}
\label{eq:cp-classifier}
    C(x) = \frac{f_+(x)}{f_+(x)+f_-(x)} 
    \; &\overset{\cp}{\longleftrightarrow} \; 
    \frac{f_-(x)}{f_-(x) + f_+(x)} = 1 - C(x) \notag \\
    \text{or} \qquad 
    C(x) = \sigma\left(\log\frac{f_+(x)}{f_-(x)}\right) 
    \; &\overset{\cp}{\longleftrightarrow} \; 
    \sigma\left(-\log\frac{f_+(x)}{f_-(x)}\right) = 1-C(x) \; ,
\end{align}
in terms of the usual sigmoid required for a probabilistic classifier output. The learned optimal $\cp$-observable is the argument of the classifier-specific sigmoid,
\begin{align}
%\label{eq:cp-llr}
    O(x) = \sigma^{-1} \left( C(x) \right) = \log\frac{f_+(x)}{f_-(x)} \; .
\end{align}
The expanded L-GATr-slim splits any phase space distribution into a scalar and a pseudoscalar contribution,
\begin{align}
    O(x) %= \log \frac{f_+(x)}{f_-(x)} 
    \equiv s(x) + p(x) \; .
\end{align}
During training, the learned scalar contribution $s(x)$ will converge to zero and leave us with
\begin{align}
\label{eq:lgatr-obs}
    O_\text{L-GATr}(x) = p(x) = \log \frac{f_+(x)}{f_-(x)} \; .
\end{align}
%

%%%%%%%%%%%%%%%%%%%%%%%%%%%%%%%%%%%%%%%%%%%%%%%%%%
\subsection{WBF Higgs production}
\label{subsec:cp_vbf}

We first test our approach on naive WBF Higgs production without decays
\begin{align}
  pp \to q q H \; ,
\end{align}
where we know the optimal \cp-odd observable analytically~\cite{Plehn:2001nj,Hankele:2006ma,Brehmer:2017lrt,Butter:2021rvz,Bahl:2025jtk}. We use MadGraph~\cite{Alwall:2014hca} to generate $10^6$ parton-level events for three values of the \cp-odd SMEFT coefficient
\begin{align}
  \mathcal{O}_{H\Tilde{W}} = \phi^\dagger \phi \Tilde W_{\mu\nu}^{I} W^{I\mu\nu}
  \qquad \text{and} \qquad
  \frac{c_{H\Tilde{W}}}{\Lambda} = \{ -1, 0, 1\} \times \frac{1}{\text{TeV}} \; 
\end{align}
The ordered set of 4-momenta of the Higgs boson and quarks in the final state are the network input. We split the dataset into $1.6 \cdot 10^6$ training, $2 \cdot 10^5$ validation, and $2 \cdot 10^5$ test events. We train five networks per task and quote the mean AUC, with the standard deviation over the trainings added in quadrature to the bootstrap uncertainty, as in Eq.\eqref{eq:network-uncertainty}.

%----------------------------------------
\begin{figure}[b!]
    \includegraphics[width=0.49\textwidth]{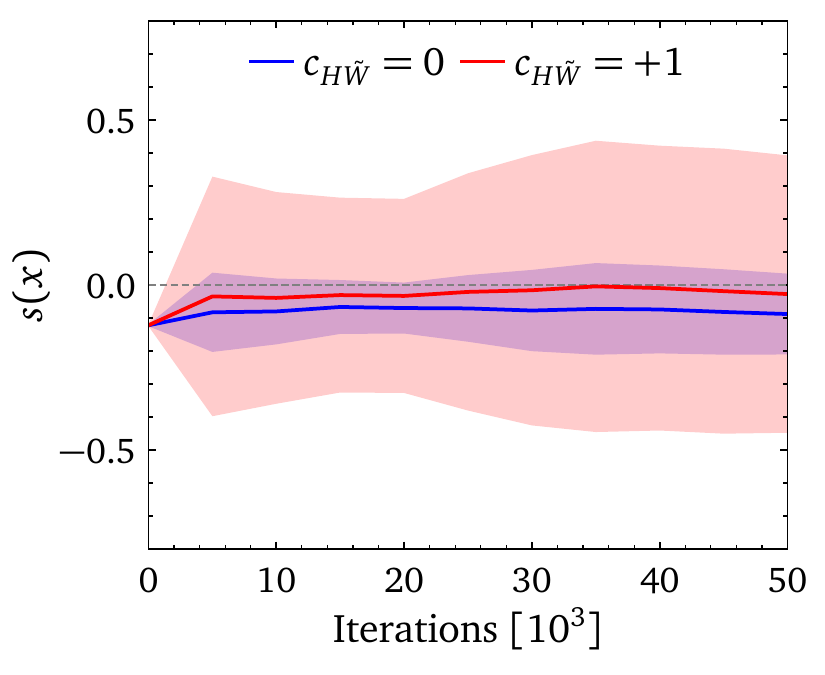}
    \includegraphics[width=0.49\textwidth]{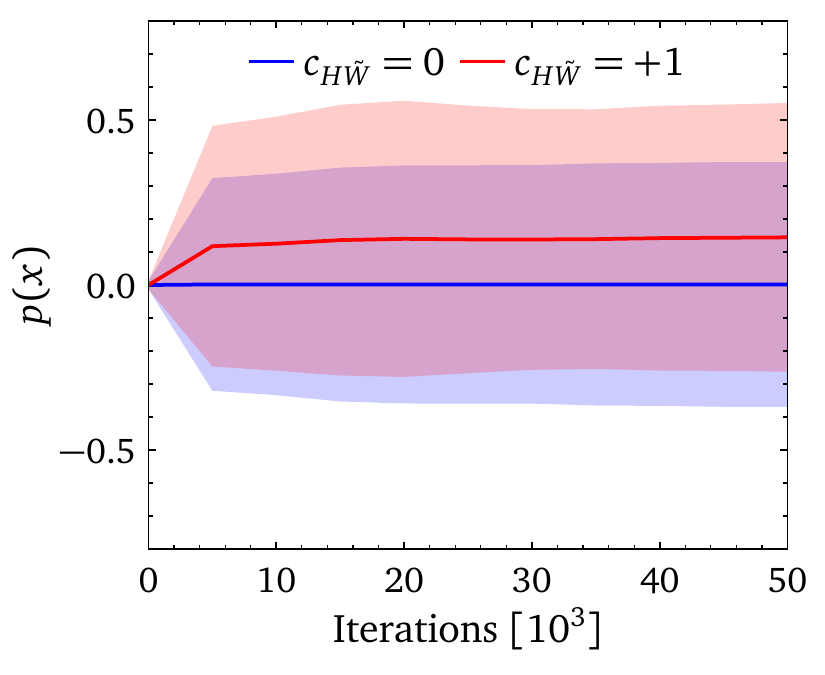}
    \caption{Mean and standard deviation of the scalar (left) and pseudoscalar outputs (right) of L-GATr-slim classifying SM events from events with $c_{H\Tilde{W}} = 1$ in WBF. We evaluate the network every 5000 iterations on the validation data and show the moving average over three validation points for one training.}
    \label{fig:lgatr_vbf_cp}
\end{figure}
%----------------------------------------

First, we train L-GATr-slim to separate events with
\begin{align}
 c_{H\Tilde{W}} = +1 
 \qquad \text{versus} \qquad 
 c_{H\Tilde{W}} = 0 \; .
\end{align}
Because the optimal \cp-observable includes a \cp-even prefactor, as shown in Eq.\eqref{eq:epsilon-cpodd}, this classification requires \cp-even and \cp-odd information. Its challenging nature is reflected in a moderate AUC of $0.607 \pm 0.001$. 

%----------------------------------------
\begin{figure}[t]
    \includegraphics[width=0.49\textwidth]{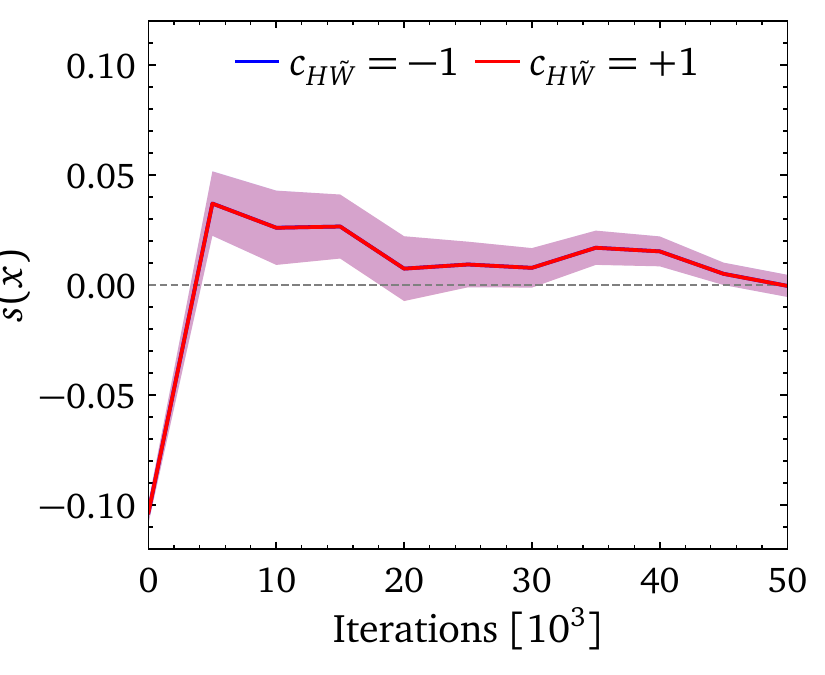}
    \includegraphics[width=0.49\textwidth]{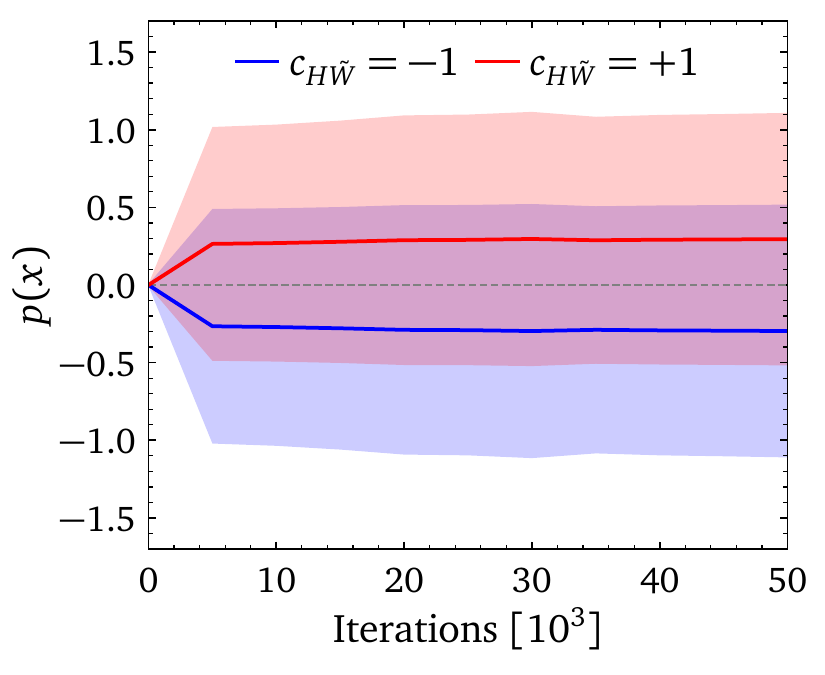}
    \caption{Mean and standard deviation of the scalar (left) and pseudoscalar outputs (right) of L-GATr-slim determining the sign of $c_{H\Tilde{W}}$ in WBF. We evaluate the network every 5000 iterations on the validation data and show the moving average over three validation points for one training.}
    \label{fig:lgatr_vbf_pseudoscalar}
\end{figure}
%----------------------------------------

We can examine the \cp structure of the classifier by looking at the scalar and pseudoscalar outputs in Fig.~\ref{fig:lgatr_vbf_cp}. We see that $s(x)$ poorly separates the SM and BSM events in the mean. The scalar separation is driven by $\abs{\mathcal{M}_\cp}^2 / \abs{\mathcal{M}_\text{SM}}^2$, which varies more for $c_{H\Tilde{W}} = +1$ than for the SM. For the pseudoscalar output $p(x)$, the SM events without \cp violation remain zero on average, but the events with \cp-odd interference give a positive output. This way, L-GATr-slim disentangles the \cp property of the underlying density.

Next, we train L-GATr-slim to differentiate between positive and negative interference,
\begin{align}
 c_{H\Tilde{W}} = +1 
 \qquad \text{versus} \qquad 
 c_{H\Tilde{W}} = -1 \; .
\end{align}
This classification only uses \cp-odd information, so the pseudoscalar L-GATr-slim output is optimal, as discussed in Eq.\eqref{eq:lgatr-obs}. It is a simpler task than before, and we achieve an AUC of $0.691 \pm 0.002$. In Fig.~\ref{fig:lgatr_vbf_pseudoscalar} we confirm that the separation relies only on the pseudoscalar output. The scalar output $s(x)$ converges towards zero for positive and negative interference events, indicating that the network learns that the data contains no separable \cp-even information. In contrast, the pseudoscalar output $p(x)$ splits the dataset into the components with positive and negative interference. 

%%%%%%%%%%%%%%%%%%%%%%%%%%%%%%%%%%%%%%%%%%%%%%%%%%
\subsection{Top pair production}
\label{subsec:cp_tt}

In this main results section, we show how to classify the \cp properties of top pair production at the detector level. As discussed in Sec.~\ref{sec:top_quark}, we induce \cp violation through an imaginary part of $c_{tG}$. We generate events using the SMEFTsim top model~\cite{Brivio:2020onw} with~\cite{CMS:2022quh}
\begin{align}
 \text{Im}(c_{tG}) \in \Big\{ \; -2.6, -0.1, 0, 0.1, 2.6 \; \Big\} \; .
\end{align}
Aside from the SMEFT model, the event generation follows Sec.~\ref{sec:downtagging}. 

We proceed in three steps. First, we use the L-GATr-slim down tagger from Sec.~\ref{sec:downtagging} to construct the standard \cp-odd observable as a benchmark. Second, we identify \cp violation by training L-GATr-slim to separate SM events from events with $\text{Im}(c_{tG}) > 0$. Third, we analyze the \cp violation by determining the sign of $\text{Im}(c_{tG})$ with the pseudoscalar output as a learned optimal \cp-odd observable.

%%%%%%%%%%%%%%%%%%%%%%%%%%%%%%%%%%%%%%%%%%%%%%%%%%
\subsubsection*{\cp-odd observable from down-tagging}

%----------------------------------------
\begin{figure}[t]
    \includegraphics[width=0.49\textwidth]{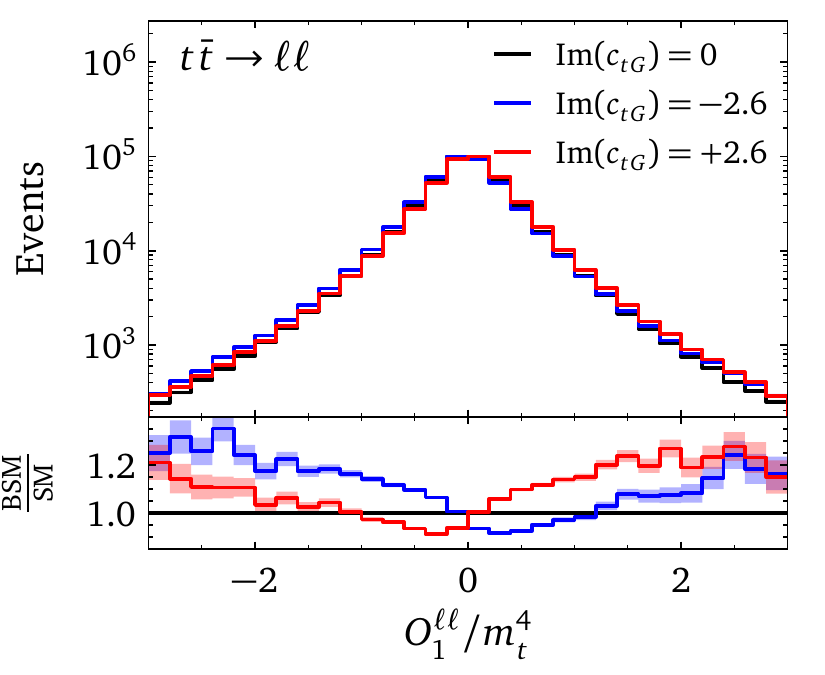}
    \includegraphics[width=0.49\textwidth]{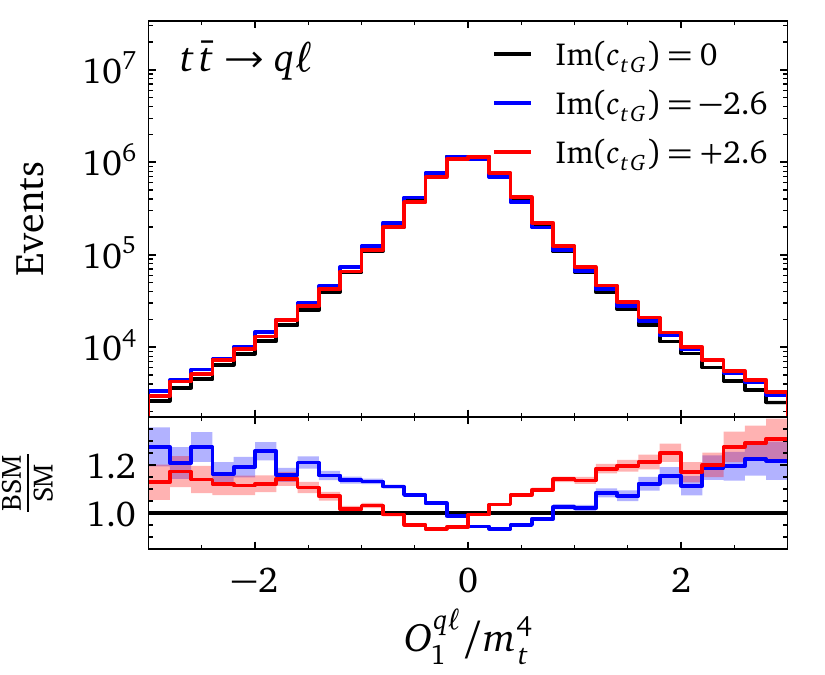}
    \caption{Distribution of $O_1$ for the SM (black), positive (red) and negative (blue) interference in the lepton-lepton (left) and lepton-hadron (right) $\ttbar$ decay channels, normalized to the expected number of events for $\mathcal{L} = 35.9\invfb$~\cite{CMS:2022quh}. The lower panels show the ratio to the SM.}
    \label{fig:O1_tt}
\end{figure}
%----------------------------------------

We consider that purely leptonic top pairs are equivalent to our lepton-hadron events with perfect down quark identification and set $\beta_\ell \equiv \beta_d = 1$. For lepton-hadron pairs, we assign the probabilistic tagging output for the down quark and adapt Eq.\eqref{eq:epsilon-semilep} to construct a 4-momentum from $\tilde k_d$ as a proxy for the true $k_d$. The corresponding \cp-odd observables are
\begin{align}
\label{cp-odd-observables}
O_1^{\ell\ell} = \epsilon_{\mu\nu\rho\sigma} k_t^\mu k_{\bar{t}}^\nu k_{\ell}^\rho k_{d}^\sigma 
\qquad \text{and} \qquad 
O_1^{q\ell} = \epsilon_{\mu\nu\rho\sigma} k_t^\mu k_{\bar{t}}^\nu k_{\ell}^\rho \tilde k_{d}^\sigma \; .
\end{align}
The asymmetries of these \cp-odd observables separate events with
\begin{align}
 \text{Im}(c_{tG}) \ne 0
 \qquad \text{versus} \qquad 
 \text{Im}(c_{tG}) = 0 \; .
\end{align}
In Fig.~\ref{fig:O1_tt} we show their distributions for different Wilson coefficients. For lepton-lepton events, the efficiently available polarization information means that differences between the three contributions are more pronounced. On the other hand, the lepton-hadron events benefit from roughly ten times more events due to the branching ratio and the acceptance. 

To illustrate the power of the \cp-odd observables, we form the asymmetry 
\begin{align}
\label{eq:asymmetry}
    A \left(\text{Im}(c_{tG})\right) &= \frac{|N^+ - N^-|}{N^+ + N^-} \notag \\
    \Rightarrow \qquad 
    \Delta_\text{stat} A \left(\text{Im}(c_{tG})\right) &=  2 \sqrt{\frac{N^+ N^-}{(N^+ + N^-)^3}} \; ,
\end{align}
where $N^\pm$ is the number of events with $O_1(\text{Im}(c_{tG})) \gtrless 0$. For lepton-lepton and lepton-hadron top decays, the complex Wilson coefficients lead to the asymmetries
\begin{align}
%\label{eq:asym_ll}
    A^{\ell\ell}_{O_1} (2.6) &= 0.0591 \pm 0.0007 
    &\qqquad 
    A^{\ell\ell}_{O_1} (0.1) &= 0.0046 \pm 0.0007 \notag \\
    A^{q\ell}_{O_1} (2.6) &= 0.0424 \pm 0.0007
    &\qqquad
    A^{q\ell}_{O_1} (0.1) &= 0.0031 \pm 0.0007 \; .
\label{eq:asym_ql}
\end{align}
The uncertainties here are statistical uncertainties of the dataset and do not correspond to expected measurements. 
%

%%%%%%%%%%%%%%%%%%%%%%%%%%%%%%%%%%%%%%%%%%%%%%%%%%
\subsubsection*{Identifying \cp violation with L-GATr-slim}

Next, we use L-GATr-slim to identify \cp violation in top pair events. In analogy to the WBF case, we train a classifier to distinguish
\begin{align}
 \text{Im}(c_{tG}) = 2.6
 \qquad \text{versus} \qquad 
 \text{Im}(c_{tG}) = 0 \; .
\end{align}
This classification requires \cp-even and \cp-odd information, so we expect L-GATr-slim to pick up on both. We train on $1.44 \cdot 10^6$ events and evaluate on $1.80 \cdot 10^5$ events. We provide further details on the network inputs and hyperparameters in App.~\ref{app:inputs-hyperparameters}.

%----------------------------------------
\begin{figure}[t]
    \includegraphics[width=0.495\textwidth]{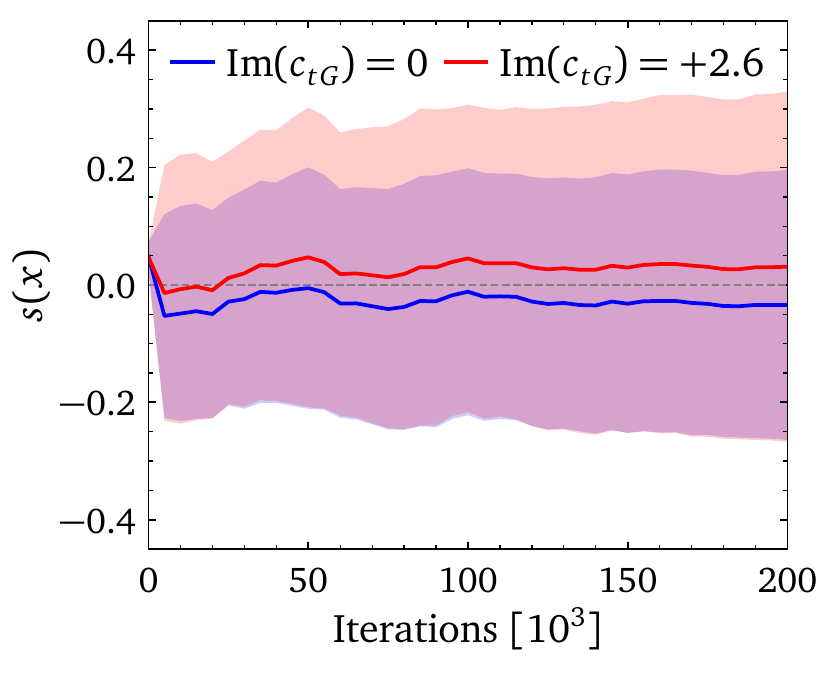}
    \includegraphics[width=0.495\textwidth]{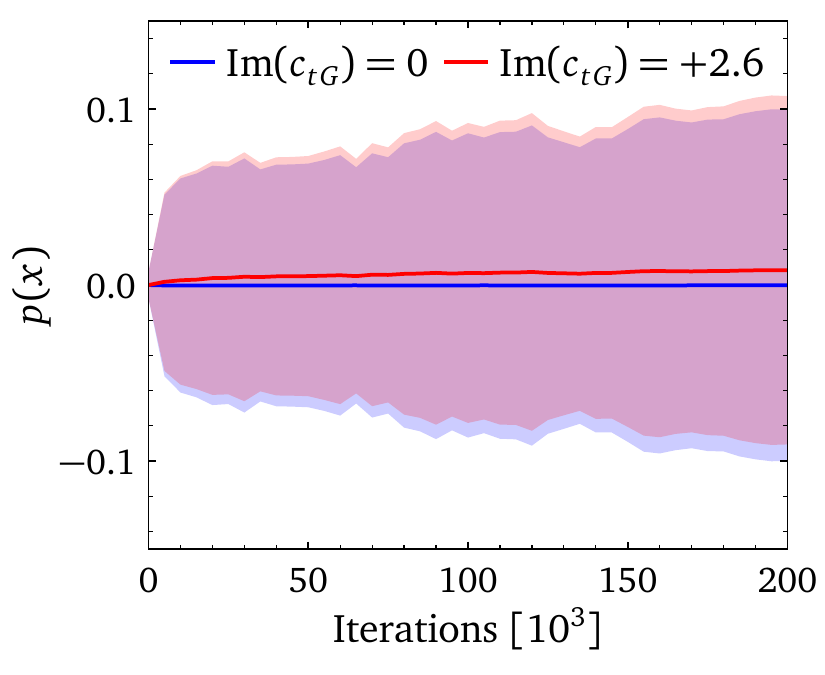}
    \caption{Mean and standard deviation of the scalar (left) and pseudoscalar outputs (right) of L-GATr-slim classifying SM events from events with $\text{Im}(c_{tG})=+2.6$. We evaluate the network every 5000 iterations on the validation data and show the moving average over five validation points for one training.}
    \label{fig:lgatr_ttsmeft_sm_channels}
\end{figure}
%----------------------------------------

In Fig.~\ref{fig:lgatr_ttsmeft_sm_channels} we show the scalar and pseudoscalar classification outputs. In combination, L-GATr-slim reaches an AUC of $0.565 \pm 0.002$ with the true down quark and $0.556 \pm 0.002$ with the tagged down quark. We observe the same behavior as in the WBF case despite the lower performance. The scalar output separates the SM and BSM events in the mean and is broader for $\text{Im}(c_{tG}) = 2.6$, where the \cp-even term $|\mathcal{M}_\text{CP}|^2$ modifies the kinematics. The pseudoscalar output averages to zero for the SM events and to a positive value for $\text{Im}(c_{tG}) = 2.6$, which includes the \cp-odd interference. Its shift is much smaller than the scalar one, so the \cp-even information leads this classification.

%%%%%%%%%%%%%%%%%%%%%%%%%%%%%%%%%%%%%%%%%%%%%%%%%%
\subsubsection*{\cp-analysis from L-GATr-slim}

%----------------------------------------
\begin{figure}[b!]
    \includegraphics[width=0.49\textwidth]{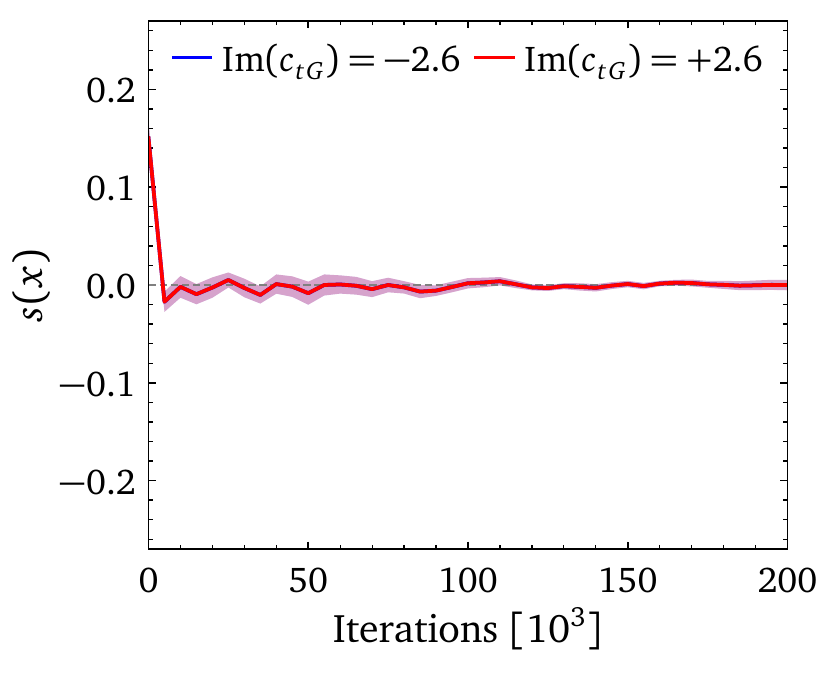}
    \includegraphics[width=0.49\textwidth]{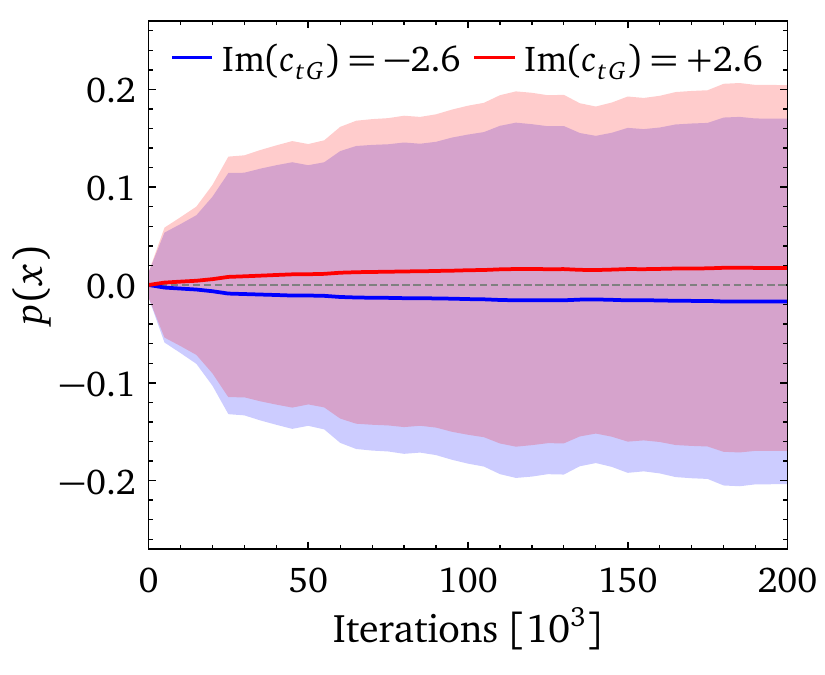}
    \caption{Mean and standard deviation of the scalar (left) and pseudoscalar outputs (right) of L-GATr-slim determining the sign of $\text{Im}(c_{tG})$. We evaluate the network every 5000 iterations on the validation data and show the moving average over five validation points for one training.}
    \label{fig:lgatr_ttsmeft_posneg_channels}
\end{figure}
%----------------------------------------

Finally, we show how we can analyze \cp violation in top pair events using the \cp-odd observable learned by L-GATr-slim. We want to determine the sign of the \cp-odd interference term
\begin{align}
 \text{Im}(c_{tG}) = 2.6
 \qquad \text{versus} \qquad 
 \text{Im}(c_{tG}) = -2.6 \; .
\end{align}
We use the same training setup as for the previous analysis, with $1.56 \cdot 10^6$ training events and $1.96 \cdot 10^5$ events for evaluation.

In Fig.~\ref{fig:lgatr_ttsmeft_posneg_channels} we show the scalar and pseudoscalar L-GATr-slim outputs. We first see that the scalar output $s(x)$ does not include any information and is moved to zero, just like in Sec.~\ref{subsec:cp_vbf}. The pseudoscalar output $p(x)$ captures the \cp-odd information and shows a clear separation of the two signs early in the training. We reach an AUC of $0.551 \pm 0.002$.

\begin{figure}[t]
    \includegraphics[width=0.495\textwidth]{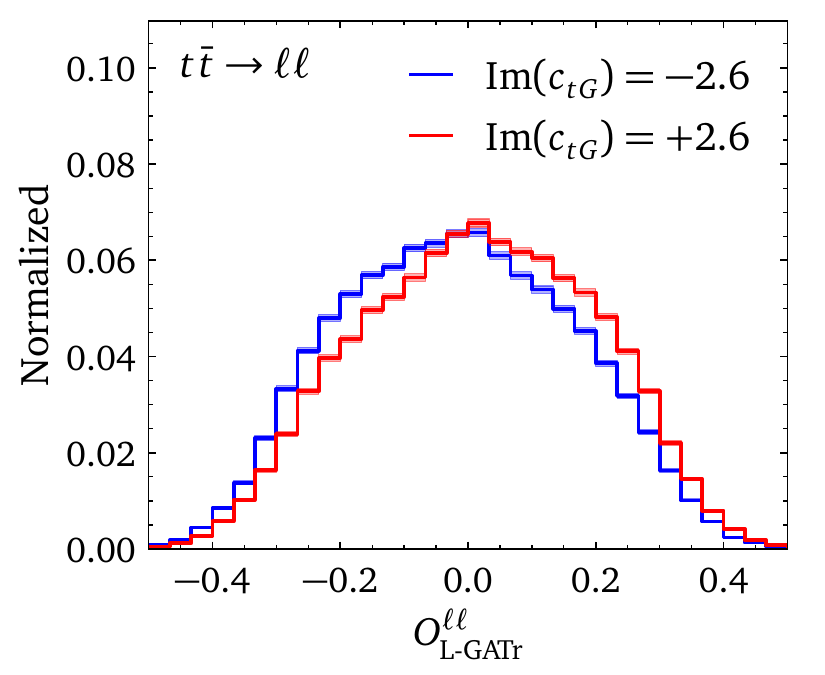}
    \includegraphics[width=0.495\textwidth]{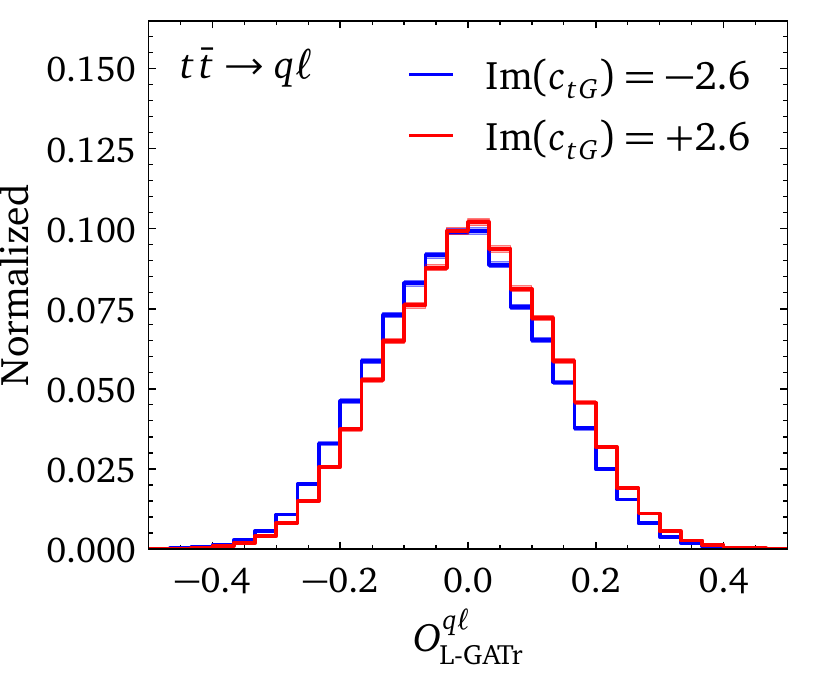}
    \caption{Pseudoscalar output of the L-GATr-slim classifier extracting the sign of $\text{Im}(c_{tG})$ for lepton-lepton (left) and lepton-hadron (right) decays.}
    \label{fig:lgatr_ttsmeft_posneg_scores}
\end{figure}

From Eq.\eqref{eq:lgatr-obs} we know that $p(x)$ is the optimal \cp-odd observable $O_\text{L-GATr}$. As mentioned above, we use the true down quark assignment for lepton-lepton top decays. For lepton-hadron decays, we define the \cp-odd observable as weighted with the two down quark assignments,
\begin{align}
    O_\text{L-GATr} 
    = p_\text{L-GATr}(d=q_\text{hard}) \; O_\text{L-GATr} \Big|_{q_\text{hard}} 
    + p_\text{L-GATr}(d=q_\text{soft}) \; O_\text{L-GATr} \Big|_{q_\text{soft}} \; .
\end{align}
In Fig.~\ref{fig:lgatr_ttsmeft_posneg_scores} we show the observable for both decay channels. We also compute the asymmetries
\begin{align}
\label{eq:lgatr-asym}
    A^{\ell\ell}_{O_{\text{L-GATr}}} (2.6) &= 0.0733 \pm 0.0023
    &\qqquad
    A^{\ell\ell}_{O_{\text{L-GATr}}} (0.1) &= 0.0045 \pm 0.0007 \notag \\
    A^{q\ell}_{O_{\text{L-GATr}}} (2.6) &= 0.0518 \pm 0.0023
    &\qqquad
    A^{q\ell}_{O_{\text{L-GATr}}} (0.1) &= 0.0028 \pm 0.0007 \; .
\end{align}
We pick the networks with the best validation loss out of five runs and only include the statistical uncertainty, as computed for $O_1$. We compute the asymmetries for $\text{Im}(c_{tG}) = 0.1$ on the full dataset, because the network is not trained on it, whereas for $\text{Im}(c_{tG}) = 2.6$ we only use the test data, hence the larger statistical error. We find that for $\text{Im}(c_{tG}) = 2.6$ the learned observables from L-GATr-slim substantially improve over their $O_1$ counterparts in Eq.\eqref{eq:asym_ql}, while they agree within the statistical uncertainties for $\text{Im}(c_{tG}) = 0.1$.

%----------------------------------------
%----------------------------------------

%%%%%%%%%%%%%%%%%%%%%%%%%%%%%%%%%%%%%%%%%%%%%%%%%%
\section{Outlook}
\label{sec:conclusion}

Measuring the polarization of tops and correlating the polarizations of top pairs provide us with a powerful probe of BSM physics. In this paper, we have shown how a Lorentz-equivariant transformer significantly improves top polarimetry and searches for \cp violation in top pair production. To this end, we have equipped L-GATr-slim with a pseudoscalar channel. This architecture splits scalar and pseudoscalar information and allows us to learn optimal \cp-odd observables. 

First, we have reconstructed the polarization of hadronically decaying tops by tagging the down-type quark with a calibrated probability output. The resulting spin-analyzing power of $0.721$ at detector level significantly outperforms all existing methods and can be increased by using the most suitable $20\%$ top decays. 

We have applied L-GATr-slim to the toy problem of extracting \cp violation from parton-level WBF events. Distinguishing SM events from events with dimension-6 \cp violation requires scalar and pseudoscalar outputs, the pseudoscalar indicating that the signal indeed violates \cp. Determining the sign of the \cp-violating Wilson coefficient only involves the new pseudoscalar L-GATr output.

Finally, we have established our new L-GATr-based methods for \cp violation from a complex dimension-6 Wilson coefficient in top pair production, including detector simulation. First, we used the down-type tagger to construct a \cp-odd observable. Alternatively, we employed the scalar and pseudoscalar multivector elements to learn an optimal \cp-sensitive observable to extract the effect of the complex dimension-6 operator. Finally, we have shown how to determine the sign of this Wilson coefficient optimally, just using the pseudoscalar L-GATr-slim output. 

Our applications show that equivariant transformers based on a geometric algebra give us access to relevant information for BSM physics searches at the LHC through output channels with well-defined Lorentz representations.

%%%%%%%%%%%%%%%%%%%%%%%%%%%%%%%%%%%%%%%%%%%%%%%
\subsection*{Code availability}

The \href{https://github.com/heidelberg-hepml/top-polarization}{code for this paper} is available as part of the public Heidelberg HEP-ML code and tutorial library. The L-GATr-slim architecture and its pseudoscalar extension are part of the \href{https://github.com/heidelberg-hepml/lgatr/tree/slim-pseudo}{L-GATr repository}. The datasets are available upon request.

%%%%%%%%%%%%%%%%%%%%%%%%%%%%%%%%%%%%%%%%%%%%%%%%%%%
\section*{Acknowledgements}

We thank Jonas Spinner for helpful discussions about the extended L-GATr-slim architecture. This project has received funding from the European Union’s Horizon Europe research and innovation programme under the Marie Sklodowska-Curie grant agreement No~101168829, \textsl{Challenging AI with Challenges from Physics: How to solve fundamental problems in Physics by AI and vice versa} (AIPHY). MM acknowledges funding from the German Research Foundation (DFG) through project number 580386328 and through Germany’s Excellence Strategy – EXC-2123 QuantumFrontiers – 390837967. TP is supported by the Deutsche Forschungsgemeinschaft (DFG, German Research Foundation) under grant 396021762 -- TRR 257 \textsl{Particle Physics Phenomenology after the Higgs Discovery}. Computations were performed on the LUIS computing cluster at Leibniz University Hannover funded by the Deutsche Forschungsgemeinschaft (DFG, German Research Foundation) – Projektnummern 563020064, INST 187/742-1 FUGG, and INST 187/592-1 FUGG. The authors would like to thank the state of Baden-Württemberg for its support of bwHPC and the German Research Foundation (DFG) for funding under "Project number 455622343" (bwForCluster NEMO 2).

\clearpage
%%%%%%%%%%%%%%%%%%%%%%%%%%%%%%%%%%%%%%%%%%%%%%%%%%
\appendix
\section{Network inputs and hyperparameters}
\label{app:inputs-hyperparameters}

Each token in the sequence given as input to the network contains vector, scalar and pseudoscalar information. The different inputs are shown in Table~\ref{tab:network_inputs}. The structure of the jet is given by a one-hot encoding of the fat jets and subjets, passed as scalar inputs. The particle ID is also given as the standard 6-dimensional one-hot encoding. The additional kinematic features passed as scalars and pseudoscalars are computed in the lab frame. We refer to App.~\ref{app:symmetry-group} for more details on the choice of scalars and pseudoscalars. We also include the symmetry-breaking vectors from App.~\ref{app:symmetry-group} as additional tokens, with their scalars and pseudoscalars set to 0.

The light subjets are ordered by their energy in the hadronic top rest frame. For the \ttbar classification, we swap them such that the (predicted or true) down-type subjet is the first one, and the network can identify it with the one-hot encoding. For the down-tagging task, the leptonic top and the lepton are not included, and the \cp symmetry is not needed, so the pseudoscalars are passed as scalars.

%----------------------------------------
\begin{table}[b!]
\centering
\setlength{\tabcolsep}{4pt}
\begin{small} \begin{tabular}{llccccccccc}
\toprule
& & top jet & $b$ subjet & $q_\text{hard}$ & $q_\text{soft}$ & $\cdots$ & constituent & $\cdots$ & leptonic top & lepton \\
\midrule
\multirow{17}{20pt}{$s$}
& hadronic top          & 1 & 0 & 0 & 0 & & 0 & & 0 & 0 \\
& subjet                & 0 & 1 & 1 & 1 & & 0 & & 0 & 0 \\
& in $b$ subjet         & 0 & 1 & 0 & 0 & & 0 & & 0 & 0 \\
& in $q_\text{hard}$ subjet         & 0 & 0 & 1 & 0 & $\cdots$ & 1 & $\cdots$ & 0 & 0 \\
& in $q_\text{soft}$ subjet         & 0 & 0 & 0 & 1 & & 0 & & 0 & 0 \\
& leptonic top          & 0 & 0 & 0 & 0 & & 0 & & 1 & 0 \\
& lepton                & 0 & 0 & 0 & 0 & & 0 & & 0 & 1 \\
\cmidrule{2-11}
& electron              & 0 & 0 & 0 & 0 & & 0 & & 0 & 0 \\
& muon                  & 0 & 0 & 0 & 0 & & 0 & & 0 & 0 \\
& photon                & 0 & 0 & 0 & 0 & $\cdots$ & 0 & $\cdots$ & 0 & 0 \\
& charged hadron        & 0 & 0 & 0 & 0 & & 1 & & 0 & 0 \\
& neutral hadron        & 0 & 0 & 0 & 0 & & 0 & & 0 & 0 \\
\cmidrule{2-11}
& $\log p_T$            & \multicolumn{9}{c}{$\log p_{T,i}$} \\
& $\log E$              & \multicolumn{9}{c}{$\log E_i$} \\
& $\Delta \log p_T$     & \multicolumn{9}{c}{$\log p_{T,i} - \log p_{T,J}$} \\
& $\Delta \log E$       & \multicolumn{9}{c}{$\log E_i - \log E_J$} \\
& $\abs{\Delta \phi}$         & \multicolumn{9}{c}{$\abs{\phi_i - \phi_J}$} \\
& $\abs{\Delta \eta}$         & \multicolumn{9}{c}{$\abs{\eta_i - \eta_J}$} \\
& $\Delta R$            & \multicolumn{9}{c}{$\sqrt{\Delta\phi^2 + \Delta\eta^2}$} \\
\midrule
\multirow{4}{*}{$v$}
& \multirow{4}{*}{4-momentum} & \multicolumn{9}{c}{$E_i$} \\
& & \multicolumn{9}{c}{$p_{x,i}$} \\
& & \multicolumn{9}{c}{$p_{y,i}$} \\
& & \multicolumn{9}{c}{$p_{z,i}$} \\
\midrule
\multirow{2}{*}{$p$}
& charge                & 0 & 0 & 0 & 0 & $\cdots$ & $+1$ & $\cdots$ & 0 & $Q_\ell$ \\
& $\Delta \phi \cdot \Delta \eta$         & \multicolumn{9}{c}{$(\phi_i - \phi_J)(\eta_i - \eta_J)$} \\
\bottomrule
\end{tabular} \end{small}
\caption{Scalar, vector, and pseudoscalar inputs of L-GATr-slim for each token of a semileptonic \ttbar event. The constituent column shows an example of a positively charged hadron in the $q_\text{hard}$ subjet. $J$ labels the top jet that the token belongs to.}
\label{tab:network_inputs}
\end{table}
%----------------------------------------

All networks use the same backbone and training setup, listed in Tab.~\ref{tab:hyperparameters}. Only the pseudoscalar channels and the number of training iterations differ between the tasks. We use the default hyperparameters of \cite{Favaro:2026amj} for tagging with L-GATr-slim and did not perform a dedicated optimization.

%----------------------------------------
\begin{table}[t]
\centering
\begin{small} \begin{tabular}{l >{\centering\arraybackslash}p{0.2\linewidth}>{\centering\arraybackslash}p{0.2\linewidth}>{\centering\arraybackslash}p{0.2\linewidth}}
\toprule
 & Down Tagging & WBF & \ttbar \\
\midrule
Blocks, heads & \multicolumn{3}{c}{12, 8} \\
MLP ratio & \multicolumn{3}{c}{4} \\
Channels $(s, v, p)$ & $(96, 32, 0)$ & $(96, 32, 32)$ & $(96, 32, 32)$\\
\midrule
Optimizer & \multicolumn{3}{c}{Lion} \\
Learning rate & \multicolumn{3}{c}{$3 \cdot 10^{-5}$} \\
LR scheduler & \multicolumn{3}{c}{Cosine annealing} \\
Weight decay & \multicolumn{3}{c}{2} \\
Batch size & \multicolumn{3}{c}{128} \\
Iterations & $5 \cdot 10^5$ & $5 \cdot 10^4$ & $2 \cdot 10^5$ \\
\bottomrule
\end{tabular} \end{small}
\caption{Training hyperparameters of L-GATr-slim for all tasks.}
\label{tab:hyperparameters}
\end{table}
%----------------------------------------

%%%%%%%%%%%%%%%%%%%%%%%%%%%%%%%%%%%%%%%%%%%%%%%%%%
\section{L-GATr-slim symmetry group}
\label{app:symmetry-group}

The symmetry group of L-GATr-slim depends on the type of inputs that are given. The network is equivariant with respect to $\text{O}(1,3)$, under the assumption that the scalar inputs are invariant and the vector inputs are covariant. If the inputs do not satisfy that property, the network is no longer equivariant: the symmetries that are broken at the input level can also be broken at the output level.

We know that detector data does not respect the full Lorentz group symmetry, so we include in the input three reference vectors that are identical in every event to break those symmetries: $t=(1,0,0,0)$ and $n_\pm=(1,0,0,\pm1)$, where $n_\pm$ defines the beam axis in the $z$ direction. Because those vectors are fixed and do not transform, they are only covariant under the following transformations:
\begin{itemize}
    \item rotations around the beam axis $R_z(\theta)$
    \item rotations by $\pi$ around any transverse axis, for example the beam swap $R_x(\pi)$
    \item mirrors through planes containing the beam axis, for example $\sigma_y=\mathcal{P}\,R_y(\pi)$
    \item rotations around the beam axis combined with the flip $\sigma_z=\mathcal{P}R_z(\pi)$
\end{itemize}
Therefore, the L-GATr-slim network with those symmetry-breaking vector inputs is only equivariant under
\begin{align}
    \text{SO}(2)_z\rtimes \{1,R_x(\pi)\} \times \{1,\mathcal{P}\}
\end{align}
where the rotation $R_x(\pi)$ can be replaced with a rotation around any other axis perpendicular to $z$.

However, we are interested here in the \cp symmetry instead of $\mathcal{P}$. The $\mathcal{C}$ transformation only acts on the charges and not on the Minkowski space $\mathbb{R}^{1,3}$, hence it commutes with all elements of $\text{O}(1,3)$. Therefore, we can write the remaining symmetry group as
\begin{align}
    \text{SO}(2)_z\rtimes \{1,R_x(\pi)\} \times\{1,\cp\} \; .
\end{align}
The same issue arises with the scalar inputs: the usual scalars given to a jet tagger, as those in Eq.\eqref{eq:scalar-inputs}, are only invariant under $\text{SO}(2)_z$ and not under $\text{O}(1,3)$. Therefore, the standard L-GATr-slim tagger is $\text{SO}(2)_z$-equivariant, which was sufficient for previous applications.

In our case, we have to separate the scalars into four categories based on their transformations under $R_x(\pi)$ and \cp. The transformations of the relevant inputs are shown in Tab.~\ref{tab:symmetry-reps}. We build our architecture to handle the $\cp$ symmetry and neglect the $R_x(\pi)$ symmetry. Therefore, we only include scalars that are invariant under $R_x(\pi)$. We find that it makes no difference in our results to break this symmetry. It could also be built into the network by adding additional channels in the same way we added the pseudoscalar channels, but that presents no benefits here.

%----------------------------------------
\begin{table}[t]
\centering
\begin{tabular}{ll|cc}
\toprule
$R_x(\pi)$ & \cp & Example & L-GATr-slim input \\
\midrule
even & even & $\abs{\Delta\phi}$, $\abs{\Delta\eta}$, $\Delta R$ & $s$ \\
even & odd & $(\Delta\phi\cdot\Delta\eta)$, charge & $p$ \\
odd & even & $\Delta\phi$ & / \\
odd & odd & $\Delta\eta$ & / \\
\bottomrule
\end{tabular}
\caption{Transformation of 1-dimensional inputs under $\{1,R_x(\pi)\} \times\{1,\cp\}$ and how they are used in the L-GATr-slim input channels.}
\label{tab:symmetry-reps}
\end{table}
%----------------------------------------

\clearpage
%%%%%%%%%%%%%%%%%%%%%%%%%%%%%%%%%%%%%%%%%%%%%%%%%%
\bibliographystyle{tepml}
\bibliography{tilman,refs}

@article{Hankele:2006ma,
    author = "Hankele, V. and Klamke, G. and Zeppenfeld, D. and Figy, T.",
    title = "{Anomalous Higgs boson couplings in vector boson fusion at the CERN LHC}",
    eprint = "hep-ph/0609075",
    archivePrefix = "arXiv",
    reportNumber = "KA-TP-09-2006",
    doi = "10.1103/PhysRevD.74.095001",
    journal = "Phys. Rev. D",
    volume = "74",
    pages = "095001",
    year = "2006"
}

@article{Leigh:2024ked,
    author = "Leigh, Matthew and Klein, Samuel and Charton, Fran{\c{c}}ois and Golling, Tobias and Heinrich, Lukas and Kagan, Michael and Ochoa, In{\^e}s and Osadchy, Margarita",
    title = "{Is Tokenization Needed for Masked Particle Modelling?}",
    eprint = "2409.12589",
    archivePrefix = "arXiv",
    primaryClass = "hep-ph",
    month = "9",
    year = "2024"
}

@article{Vigl:2026bah,
    author = "Vigl, Matthias and Pond, Nikita and Barr, Jackson and Froch, Alexander and Guest, Dan and Hartman, Nicole and Kagan, Michael and Heinrich, Lukas",
    title = "{How to scale your HEP ML models: A recipe for robust architecture comparisons at scale}",
    eprint = "2610.06784",
    archivePrefix = "arXiv",
    primaryClass = "hep-ex",
    month = "10",
    year = "2026"
}

@article{Mikuni:2021pou,
    author = "Mikuni, Vinicius and Canelli, Florencia",
    title = "{Point cloud transformers applied to collider physics}",
    eprint = "2102.05073",
    archivePrefix = "arXiv",
    primaryClass = "physics.data-an",
    doi = "10.1088/2632-2153/ac07f6",
    journal = "Mach. Learn. Sci. Tech.",
    volume = "2",
    number = "3",
    pages = "035027",
    year = "2021"
}

@article{Qiu:2022xvr,
    author = "Qiu, Shikai and Han, Shuo and Ju, Xiangyang and Nachman, Benjamin and Wang, Haichen",
    title = "{Holistic approach to predicting top quark kinematic properties with the covariant particle transformer}",
    eprint = "2203.05687",
    archivePrefix = "arXiv",
    primaryClass = "hep-ph",
    doi = "10.1103/PhysRevD.107.114029",
    journal = "Phys. Rev. D",
    volume = "107",
    number = "11",
    pages = "114029",
    year = "2023"
}

@article{DiBello:2022iwf,
    author = "Di Bello, Francesco Armando and others",
    title = "{Reconstructing particles in jets using set transformer and hypergraph prediction networks}",
    eprint = "2212.01328",
    archivePrefix = "arXiv",
    primaryClass = "hep-ex",
    doi = "10.1140/epjc/s10052-023-11677-7",
    journal = "Eur. Phys. J. C",
    volume = "83",
    number = "7",
    pages = "596",
    year = "2023"
}

@article{Finke:2023veq,
    author = {Finke, Thorben and Kr{\"a}mer, Michael and M{\"u}ck, Alexander and T{\"o}nshoff, Jan},
    title = "{Learning the language of QCD jets with transformers}",
    eprint = "2303.07364",
    archivePrefix = "arXiv",
    primaryClass = "hep-ph",
    doi = "10.1007/JHEP06(2023)184",
    journal = "JHEP",
    volume = "06",
    pages = "184",
    year = "2023"
}

@article{Qu:2022mxj,
    author = "Qu, Huilin and Li, Congqiao and Qian, Sitian",
    title = "{Particle Transformer for Jet Tagging}",
    eprint = "2202.03772",
    archivePrefix = "arXiv",
    primaryClass = "hep-ph",
    month = "2",
    year = "2022"
}

@article{Han:2009ra,
    author = "Han, Tao and Li, Yingchuan",
    title = "{Genuine CP-odd Observables at the LHC}",
    eprint = "0911.2933",
    archivePrefix = "arXiv",
    primaryClass = "hep-ph",
    reportNumber = "MADPH-09-1549, NPAC-09-15",
    doi = "10.1016/j.physletb.2009.12.047",
    journal = "Phys. Lett. B",
    volume = "683",
    pages = "278--281",
    year = "2010"
}

@article{Atwood:2000tu,
    author = "Atwood, David and Bar-Shalom, Shaouly and Eilam, Gad and Soni, Amarjit",
    title = "{CP violation in top physics}",
    eprint = "hep-ph/0006032",
    archivePrefix = "arXiv",
    reportNumber = "AMES-HET-00-03, BNL-HET-00-7, ROME1-1294-00",
    doi = "10.1016/S0370-1573(00)00112-5",
    journal = "Phys. Rept.",
    volume = "347",
    pages = "1--222",
    year = "2001"
}

@article{Krause:2026ayh,
    author = "Krause, Claudius and Winterhalder, Ramon and Feickert, Matthew and Nachman, Benjamin",
    title = "{The Living Guide of Machine Learning for Particle Physics}",
    eprint = "2608.09531",
    archivePrefix = "arXiv",
    primaryClass = "hep-ph",
    reportNumber = "MBI-ML-26-05, TIF-UNIMI-2026-10",
    month = "8",
    year = "2026"
}

@article{Brivio:2017vri,
    author = "Brivio, Ilaria and Trott, Michael",
    title = "{The Standard Model as an Effective Field Theory}",
    eprint = "1706.08945",
    archivePrefix = "arXiv",
    primaryClass = "hep-ph",
    doi = "10.1016/j.physrep.2018.11.002",
    journal = "Phys. Rept.",
    volume = "793",
    pages = "1--98",
    year = "2019"
}

@article{Bigi:1986jk,
    author = "Bigi, Ikaros I. Y. and Dokshitzer, Yuri L. and Khoze, Valery A. and Kuhn, Johann H. and Zerwas, Peter M.",
    title = "{Production and Decay Properties of Ultraheavy Quarks}",
    reportNumber = "SLAC-PUB-4021, CERN-TH-4494-86",
    doi = "10.1016/0370-2693(86)91275-X",
    journal = "Phys. Lett. B",
    volume = "181",
    pages = "157--163",
    year = "1986"
}

@article{Bierlich:2022pfr,
    author = "Bierlich, Christian and others",
    title = "{A comprehensive guide to the physics and usage of PYTHIA 8.3}",
    eprint = "2203.11601",
    archivePrefix = "arXiv",
    primaryClass = "hep-ph",
    reportNumber = "LU-TP 22-16, MCNET-22-04, FERMILAB-PUB-22-227-SCD",
    doi = "10.21468/SciPostPhysCodeb.8",
    journal = "SciPost Phys. Codeb.",
    volume = "2022",
    pages = "8",
    year = "2022"
}

@article{Bahl:2021dnc,
	author       = {Bahl, Henning and Brass, Simon},
	title        = {{Constraining CP-violation in the Higgs-top-quark interaction using machine-learning-based inference}},
	year         = 2021,
	month        = 10,
	journal      = {JHEP},
	volume       = {03},
	pages        = {017},
	doi          = {10.1007/JHEP03(2022)017},
	eprint       = {2110.10177},
	archiveprefix = {arXiv},
	primaryclass = {hep-ph},
	reportnumber = {DESY 21-160}
}

@article{Barman:2021yfh,
	author       = {Barman, Rahool Kumar and Gon\c{c}alves, Dorival and Kling, Felix},
	title        = {{Machine Learning the Higgs-Top CP Phase}},
	year         = 2021,
	month        = 10,
	journal      = {Phys. Rev. D},
	volume       = 105,
	number       = 3,
	pages        = {035023},
	doi          = {10.1103/PhysRevD.105.035023},
	eprint       = {2110.07635},
	archiveprefix = {arXiv},
	primaryclass = {hep-ph},
	reportnumber = {DESY 21-161}
}

@article{Bortolato:2020zcg,
	author       = {Bortolato, Bla\v{z} and Kamenik, Jernej F. and Ko\v{s}nik, Nejc and Smolkovi\v{c}, Aleks},
	title        = {{Optimized probes of $CP$ -odd effects in the $t \bar t h$ process at hadron colliders}},
	year         = 2021,
	journal      = {Nucl. Phys. B},
	volume       = 964,
	pages        = 115328,
	doi          = {10.1016/j.nuclphysb.2021.115328},
	eprint       = {2006.13110},
	archiveprefix = {arXiv},
	primaryclass = {hep-ph}
}

@article{Qu:2019gqs,
	author       = {Qu, Huilin and Gouskos, Loukas},
	title        = {{ParticleNet: Jet Tagging via Particle Clouds}},
	year         = 2020,
	journal      = {Phys. Rev. D},
	volume       = 101,
	number       = 5,
	pages        = {056019},
	doi          = {10.1103/PhysRevD.101.056019},
	eprint       = {1902.08570},
	archiveprefix = {arXiv},
	primaryclass = {hep-ph}
}

@article{Cranmer:2015bka,
	author       = {Cranmer, Kyle and Pavez, Juan and Louppe, Gilles},
	title        = {{Approximating Likelihood Ratios with Calibrated Discriminative  Classifiers}},
	year         = 2015,
	month        = 6,
	eprint       = {1506.02169},
	archiveprefix = {arXiv},
	primaryclass = {stat.AP},
	slaccitation = {%%CITATION = ARXIV:1506.02169;%%}
}

@article{Alwall:2014hca,
	author       = {Alwall, J. and Frederix, R. and Frixione, S. and Hirschi, V. and Maltoni, F. and Mattelaer, O. and Shao, H. -S. and Stelzer, T. and Torrielli, P. and Zaro, M.},
	title        = {{The automated computation of tree-level and next-to-leading order differential cross sections, and their matching to parton shower simulations}},
	year         = 2014,
	journal      = {JHEP},
	volume       = {07},
	pages        = {079},
	doi          = {10.1007/JHEP07(2014)079},
	eprint       = {1405.0301},
	archiveprefix = {arXiv},
	primaryclass = {hep-ph},
	reportnumber = {CERN-PH-TH-2014-064, CP3-14-18, LPN14-066, MCNET-14-09, ZU-TH-14-14}
}

@article{deFavereau:2013fsa,
	author       = {de Favereau, J. and Delaere, C. and Demin, P. and Giammanco, A. and Lemaître, V. and Mertens, A. and Selvaggi, M.},
	title        = {{DELPHES 3, A modular framework for fast simulation of a generic collider experiment}},
	year         = 2014,
	journal      = {JHEP},
	volume       = {02},
	pages        = {057},
	doi          = {10.1007/JHEP02(2014)057},
	collaboration = {DELPHES 3},
	eprint       = {1307.6346},
	archiveprefix = {arXiv},
	primaryclass = {hep-ex},
	slaccitation = {%%CITATION = ARXIV:1307.6346;%%}
}

@article{Cacciari:2011ma,
	author       = {Cacciari, Matteo and Salam, Gavin P. and Soyez, Gregory},
	title        = {{FastJet User Manual}},
	year         = 2012,
	journal      = {Eur. Phys. J. C},
	volume       = 72,
	pages        = 1896,
	doi          = {10.1140/epjc/s10052-012-1896-2},
	eprint       = {1111.6097},
	archiveprefix = {arXiv},
	primaryclass = {hep-ph},
	reportnumber = {CERN-PH-TH-2011-297}
}

@article{Dong:2024xsg,
    author = "Dong, Zhongtian and Gon{\c{c}}alves, Dorival and Kong, Kyoungchul and Larkoski, Andrew J. and Navarro, Alberto",
    title = "{Hadronic top quark polarimetry with ParticleNet}",
    eprint = "2407.01663",
    archivePrefix = "arXiv",
    primaryClass = "hep-ph",
    doi = "10.1016/j.physletb.2025.139314",
    journal = "Phys. Lett. B",
    volume = "862",
    pages = "139314",
    year = "2025"
}

@article{Dong:2024xsb,
    author = "Dong, Zhongtian and Gon{\c{c}}alves, Dorival and Kong, Kyoungchul and Larkoski, Andrew J. and Navarro, Alberto",
    title = "{Analytical insights on hadronic top quark polarimetry}",
    eprint = "2407.07147",
    archivePrefix = "arXiv",
    primaryClass = "hep-ph",
    doi = "10.1007/JHEP02(2025)117",
    journal = "JHEP",
    volume = "02",
    pages = "117",
    year = "2025"
}

@article{ATLAS:2014abv,
    author = "Aad, Georges and others",
    collaboration = "ATLAS",
    title = "{Measurement of Spin Correlation in Top-Antitop Quark Events and Search for Top Squark Pair Production in pp Collisions at $\sqrt{s}=8$ TeV Using the ATLAS Detector}",
    eprint = "1412.4742",
    archivePrefix = "arXiv",
    primaryClass = "hep-ex",
    reportNumber = "CERN-PH-EP-2014-257",
    doi = "10.1103/PhysRevLett.114.142001",
    journal = "Phys. Rev. Lett.",
    volume = "114",
    number = "14",
    pages = "142001",
    year = "2015"
}

@article{ATLAS:2014aus,
    author = "Aad, Georges and others",
    collaboration = "ATLAS",
    title = "{Measurements of spin correlation in top-antitop quark events from proton-proton collisions at $\sqrt{s}=7$ TeV using the ATLAS detector}",
    eprint = "1407.4314",
    archivePrefix = "arXiv",
    primaryClass = "hep-ex",
    reportNumber = "CERN-PH-EP-2014-116",
    doi = "10.1103/PhysRevD.90.112016",
    journal = "Phys. Rev. D",
    volume = "90",
    number = "11",
    pages = "112016",
    year = "2014"
}

@article{CMS:2013roq,
    author = "Chatrchyan, Serguei and others",
    collaboration = "CMS",
    title = "{Measurements of $t\bar{t}$ Spin Correlations and Top-Quark Polarization Using Dilepton Final States in $pp$ Collisions at $\sqrt{s}$ = 7 TeV}",
    eprint = "1311.3924",
    archivePrefix = "arXiv",
    primaryClass = "hep-ex",
    reportNumber = "CMS-TOP-13-003, CERN-PH-EP-2013-211",
    doi = "10.1103/PhysRevLett.112.182001",
    journal = "Phys. Rev. Lett.",
    volume = "112",
    number = "18",
    pages = "182001",
    year = "2014"
}

@article{ATLAS:2015igw,
    author = "Aad, Georges and others",
    collaboration = "ATLAS",
    title = "{Measurement of the correlations between the polar angles of leptons from top quark decays in the helicity basis at $\sqrt{s}=7$TeV using the ATLAS detector}",
    eprint = "1510.07478",
    archivePrefix = "arXiv",
    primaryClass = "hep-ex",
    reportNumber = "CERN-PH-EP-2015-241",
    doi = "10.1103/PhysRevD.93.012002",
    journal = "Phys. Rev. D",
    volume = "93",
    number = "1",
    pages = "012002",
    year = "2016"
}

@article{CMS:2016piu,
    author = "Khachatryan, V. and others",
    collaboration = "CMS",
    title = "{Measurements of t t-bar spin correlations and top quark polarization using dilepton final states in pp collisions at sqrt(s) = 8 TeV}",
    eprint = "1601.01107",
    archivePrefix = "arXiv",
    primaryClass = "hep-ex",
    reportNumber = "CMS-TOP-14-023, CERN-PH-EP-2015-333",
    doi = "10.1103/PhysRevD.93.052007",
    journal = "Phys. Rev. D",
    volume = "93",
    number = "5",
    pages = "052007",
    year = "2016"
}

@article{ATLAS:2019zrq,
    author = "Aaboud, Morad and others",
    collaboration = "ATLAS",
    title = "{Measurements of top-quark pair spin correlations in the $e\mu$ channel at $\sqrt{s} = 13$ TeV using $pp$ collisions in the ATLAS detector}",
    eprint = "1903.07570",
    archivePrefix = "arXiv",
    primaryClass = "hep-ex",
    reportNumber = "CERN-EP-2019-034",
    doi = "10.1140/epjc/s10052-020-8181-6",
    journal = "Eur. Phys. J. C",
    volume = "80",
    number = "8",
    pages = "754",
    year = "2020"
}

@article{CMS:2019nrx,
    author = "Sirunyan, Albert M and others",
    collaboration = "CMS",
    title = "{Measurement of the top quark polarization and $\mathrm{t\bar{t}}$ spin correlations using dilepton final states in proton-proton collisions at $\sqrt{s} =$ 13 TeV}",
    eprint = "1907.03729",
    archivePrefix = "arXiv",
    primaryClass = "hep-ex",
    reportNumber = "CMS-TOP-18-006, CERN-EP-2019-073, CMS-PAS-TOP-18-006",
    doi = "10.1103/PhysRevD.100.072002",
    journal = "Phys. Rev. D",
    volume = "100",
    number = "7",
    pages = "072002",
    year = "2019"
}

@article{CMS:2024zkc,
    author = "Hayrapetyan, Aram and others",
    collaboration = "CMS",
    title = "{Measurements of polarization and spin correlation and observation of entanglement in top quark pairs using lepton+jets events from proton-proton collisions at s=13{\,}{\,}TeV}",
    eprint = "2409.11067",
    archivePrefix = "arXiv",
    primaryClass = "hep-ex",
    reportNumber = "CMS-TOP-23-007, CERN-EP-2024-231",
    doi = "10.1103/PhysRevD.110.112016",
    journal = "Phys. Rev. D",
    volume = "110",
    number = "11",
    pages = "112016",
    year = "2024"
}

@article{CMS:2025brx,
    author = "Hayrapetyan, Aram and others",
    collaboration = "CMS",
    title = "{Characterization of the quantum state of top quark pairs produced in proton-proton collisions at $\sqrt{s}$ = 13 TeV using the beam and helicity bases}",
    eprint = "2512.17557",
    archivePrefix = "arXiv",
    primaryClass = "hep-ex",
    reportNumber = "CMS-TOP-25-016, CERN-EP-2025-287",
    month = "12",
    year = "2025"
}

@article{Tweedie:2014yda,
    author = "Tweedie, Brock",
    title = "{Better Hadronic Top Quark Polarimetry}",
    eprint = "1401.3021",
    archivePrefix = "arXiv",
    primaryClass = "hep-ph",
    reportNumber = "PITT-PACC-1315",
    doi = "10.1103/PhysRevD.90.094010",
    journal = "Phys. Rev. D",
    volume = "90",
    number = "9",
    pages = "094010",
    year = "2014"
}

@article{CMS:2009lxa,
    collaboration = "CMS",
    title = "{A Cambridge-Aachen (C-A) based Jet Algorithm for boosted top-jet tagging}",
    reportNumber = "CMS-PAS-JME-09-001, CMS-PAS-JME-09-001",
    month = "7",
    year = "2009"
}

@article{Lamba:2026sni,
    author = "Lamba, Priyanka and Maltoni, Fabio and Miniati, Olimpia and Vryonidou, Eleni",
    title = "{Quantum detection of CP violation in the $t\bar{t}$ system: tomography}",
    eprint = "2607.25034",
    archivePrefix = "arXiv",
    primaryClass = "hep-ph",
    month = "7",
    year = "2026"
}

@article{Lamba:2026jfm,
    author = "Lamba, Priyanka and Maltoni, Fabio and Miniati, Olimpia and Vryonidou, Eleni",
    title = "{Quantum detection of CP violation in the $t\bar{t}$ system: production}",
    eprint = "2607.25029",
    archivePrefix = "arXiv",
    primaryClass = "hep-ph",
    month = "7",
    year = "2026"
}

@article{Erdogan:2026cbe,
    author = "Erdogan, Eren and Mohan, Kirtimaan A. and Xie, Keping and Yanez, Marcel",
    title = "{Quantum Tomography of Top Quarks as a Probe of Charge-Parity Violation}",
    eprint = "2607.28761",
    archivePrefix = "arXiv",
    primaryClass = "hep-ph",
    reportNumber = "MSUHEP-26-011",
    month = "7",
    year = "2026"
}

@article{CMS:2022quh,
    author = "Tumasyan, Armen and others",
    collaboration = "CMS",
    title = "{Search for CP violating top quark couplings in pp collisions at $ \sqrt{s} $ = 13 TeV}",
    eprint = "2205.07434",
    archivePrefix = "arXiv",
    primaryClass = "hep-ex",
    reportNumber = "CMS-TOP-18-007, CERN-EP-2021-143",
    doi = "10.1007/JHEP07(2023)023",
    journal = "JHEP",
    volume = "07",
    pages = "023",
    year = "2023"
}

@inproceedings{dauphin2017language,
  title     = {Language Modeling with Gated Convolutional Networks},
  author    = {Dauphin, Yann N. and Fan, Angela and Auli, Michael and Grangier, David},
  booktitle = {Proceedings of the 34th International Conference on Machine Learning (ICML)},
  year      = {2017},
  eprint    = {1612.08083},
  archivePrefix = {arXiv},
  primaryClass = {cs.CL},
  volume    = {70},
  series    = {Proceedings of Machine Learning Research},
  publisher = {PMLR}
}

@article{Brivio:2020onw,
    author = "Brivio, Ilaria",
    title = "{SMEFTsim 3.0 {\textemdash} a practical guide}",
    eprint = "2012.11343",
    archivePrefix = "arXiv",
    primaryClass = "hep-ph",
    doi = "10.1007/JHEP04(2021)073",
    journal = "JHEP",
    volume = "04",
    pages = "073",
    year = "2021"
}

@article{Bernreuther:2017yhg,
    author = "Bernreuther, Werner and Galler, Peter and Si, Zong-Guo and Uwer, Peter",
    title = "{Production of heavy Higgs bosons and decay into top quarks at the LHC. II: Top-quark polarization and spin correlation effects}",
    eprint = "1702.06063",
    archivePrefix = "arXiv",
    primaryClass = "hep-ph",
    reportNumber = "HU-EP-17-04, TTK-17-03",
    doi = "10.1103/PhysRevD.95.095012",
    journal = "Phys. Rev. D",
    volume = "95",
    number = "9",
    pages = "095012",
    year = "2017"
}

@article{Abel:1992kz,
    author = "Abel, S. A. and Dittmar, M. and Dreiner, Herbert K.",
    title = "{Testing locality at colliders via Bell's inequality?}",
    reportNumber = "OUTP-91-40-P",
    doi = "10.1016/0370-2693(92)90071-B",
    journal = "Phys. Lett. B",
    volume = "280",
    pages = "304--312",
    year = "1992"
}

@article{Bechtle:2025ugc,
    author = "Bechtle, Philip and Breuning, Cedric and Dreiner, Herbi K. and Duhr, Claude",
    title = "{A critical appraisal of tests of locality and of entanglement versus non-entanglement at colliders}",
    eprint = "2507.15947",
    archivePrefix = "arXiv",
    primaryClass = "hep-ph",
    reportNumber = "BONN-TH-2025-23",
    month = "7",
    year = "2025"
}

@article{Abel:2025wwa,
    author = "Abel, Steven A. and Dreiner, Herbi K. and Sengupta, Rhitaja and Ubaldi, Lorenzo",
    title = "{Colliders are not testing locality via Bell's inequality nor providing an unconditional proof of entanglement}",
    eprint = "2507.15949",
    archivePrefix = "arXiv",
    primaryClass = "hep-ph",
    reportNumber = "BONN-TH-2025-22, IPPP/25/48",
    doi = "10.1007/JHEP08(2026)067",
    journal = "JHEP",
    volume = "08",
    pages = "067",
    year = "2026"
}

@article{Jezabek:1994zv,
    author = "Jezabek, M. and Kuhn, Johann H.",
    title = "{V-A tests through leptons from polarized top quarks}",
    eprint = "hep-ph/9403366",
    archivePrefix = "arXiv",
    reportNumber = "TTP-94-4",
    doi = "10.1016/0370-2693(94)90779-X",
    journal = "Phys. Lett. B",
    volume = "329",
    pages = "317--324",
    year = "1994"
}

@article{Brandenburg:2002xr,
    author = "Brandenburg, Arnd and Si, Z. G. and Uwer, P.",
    title = "{QCD corrected spin analyzing power of jets in decays of polarized top quarks}",
    eprint = "hep-ph/0205023",
    archivePrefix = "arXiv",
    reportNumber = "PITHA-02-07, DESY-02-055, TTP02-04",
    doi = "10.1016/S0370-2693(02)02098-1",
    journal = "Phys. Lett. B",
    volume = "539",
    pages = "235--241",
    year = "2002"
}

@article{Grzadkowski:2010es,
    author = "Grzadkowski, B. and Iskrzynski, M. and Misiak, M. and Rosiek, J.",
    title = "{Dimension-Six Terms in the Standard Model Lagrangian}",
    eprint = "1008.4884",
    archivePrefix = "arXiv",
    primaryClass = "hep-ph",
    reportNumber = "IFT-9-2010, TTP10-35",
    doi = "10.1007/JHEP10(2010)085",
    journal = "JHEP",
    volume = "10",
    pages = "085",
    year = "2010"
}

@article{Antipin:2008zx,
    author = "Antipin, Oleg and Valencia, G.",
    title = "{T-odd correlations from CP violating anomalous top-quark couplings revisited}",
    eprint = "0807.1295",
    archivePrefix = "arXiv",
    primaryClass = "hep-ph",
    doi = "10.1103/PhysRevD.79.013013",
    journal = "Phys. Rev. D",
    volume = "79",
    pages = "013013",
    year = "2009"
}

@article{Gupta:2009wu,
    author = "Gupta, Sudhir Kumar and Mete, Alaettin Serhan and Valencia, G.",
    title = "{CP violating anomalous top-quark couplings at the LHC}",
    eprint = "0905.1074",
    archivePrefix = "arXiv",
    primaryClass = "hep-ph",
    doi = "10.1103/PhysRevD.80.034013",
    journal = "Phys. Rev. D",
    volume = "80",
    pages = "034013",
    year = "2009"
}

@article{Hayreter:2015ryk,
    author = "Hayreter, Alper and Valencia, German",
    title = "{T-odd correlations from the top-quark chromoelectric dipole moment in lepton plus jets top-pair events}",
    eprint = "1511.01464",
    archivePrefix = "arXiv",
    primaryClass = "hep-ph",
    reportNumber = "COEPP-MN-15-11",
    doi = "10.1103/PhysRevD.93.014020",
    journal = "Phys. Rev. D",
    volume = "93",
    number = "1",
    pages = "014020",
    year = "2016"
}

@article{Bhardwaj:2021ujv,
    author = "Bhardwaj, Akanksha and Englert, Christoph and Hankache, Robert and Pilkington, Andrew D.",
    title = "{Machine-enhanced CP-asymmetries in the Higgs sector}",
    eprint = "2112.05052",
    archivePrefix = "arXiv",
    primaryClass = "hep-ph",
    doi = "10.1016/j.physletb.2022.137246",
    journal = "Phys. Lett. B",
    volume = "832",
    pages = "137246",
    year = "2022"
}

@article{Cruz:2024grk,
    author = "Cruz, Sergio S{\'a}nchez and Kolosova, Marina and Ram{\'o}n {\'A}lvarez, Clara and Petrucciani, Giovanni and Vischia, Pietro",
    title = "{Equivariant neural networks for robust CP observables}",
    eprint = "2405.13524",
    archivePrefix = "arXiv",
    primaryClass = "hep-ph",
    doi = "10.1103/PhysRevD.110.096023",
    journal = "Phys. Rev. D",
    volume = "110",
    number = "9",
    pages = "096023",
    year = "2024"
}

@article{Hall:2022bme,
    author = "Hall, Noah Clarke and Criddle, Isaac and Crossland, Archie and Englert, Christoph and Forbes, Patrick and Hankache, Robert and Pilkington, Andrew D.",
    title = "{Machine-enhanced CP-asymmetries in the electroweak sector}",
    eprint = "2209.05143",
    archivePrefix = "arXiv",
    primaryClass = "hep-ph",
    doi = "10.1103/PhysRevD.107.016008",
    journal = "Phys. Rev. D",
    volume = "107",
    number = "1",
    pages = "016008",
    year = "2023"
}

@article{Bahl:2023qwk,
    author = "Bahl, Henning and Fuchs, Elina and Hannig, Marc and Menen, Marco",
    title = "{Classifying the CP properties of the ggH coupling in H + 2j production}",
    eprint = "2309.03146",
    archivePrefix = "arXiv",
    primaryClass = "hep-ph",
    reportNumber = "CERN-TH-2023-158",
    doi = "10.21468/SciPostPhysCore.8.1.006",
    journal = "SciPost Phys. Core",
    volume = "8",
    pages = "006",
    year = "2025"
}

@article{Barrue:2023ysk,
    author = "Barru{\'e}, Ricardo and Conde-Mu{\'\i}{\~n}o, Patricia and Dao, Valerio and Santos, Rui",
    title = "{Simulation-based inference in the search for CP violation in leptonic WH production}",
    eprint = "2308.02882",
    archivePrefix = "arXiv",
    primaryClass = "hep-ph",
    doi = "10.1007/JHEP04(2024)014",
    journal = "JHEP",
    volume = "04",
    pages = "014",
    year = "2024"
}

@article{Esmail:2024gdc,
    author = "Esmail, Waleed and Hammad, A. and Jueid, Adil and Moretti, Stefano",
    title = "{Boosting probes of $ \mathcal{CP} $ violation in the top Yukawa coupling with Deep Learning}",
    eprint = "2405.16499",
    archivePrefix = "arXiv",
    primaryClass = "hep-ph",
    reportNumber = "CTPU-PTC-24-14",
    doi = "10.1007/JHEP12(2025)050",
    journal = "JHEP",
    volume = "12",
    pages = "050",
    year = "2025"
}

@article{Boto:2026gzj,
    author = {Boto, Rafael and Elyaouti, Karim and Fontes, Duarte and Gon{\c{c}}alves, Maria and M{\"u}hlleitner, Margarete and Rom{\~a}o, Jorge C. and Santos, Rui and Silva, Jo{\~a}o P.},
    title = "{Reassessing CP violation in the complex 2HDM with machine learning}",
    eprint = "2601.15227",
    archivePrefix = "arXiv",
    primaryClass = "hep-ph",
    reportNumber = "KA-TP-03-2026, P3H-26-005",
    doi = "10.1103/1pdt-tmcb",
    journal = "Phys. Rev. D",
    volume = "114",
    number = "5",
    pages = "055010",
    year = "2026"
}

@article{Perelstein:2026dbu,
    author = "Perelstein, Maxim and Shaik, Minal and Youn, Taewook",
    title = "{Probing CP-violating top-quark dipole moments with tomographic observables}",
    eprint = "2608.12463",
    archivePrefix = "arXiv",
    primaryClass = "hep-ph",
    month = "8",
    year = "2026"
}

@article{Favaro:2026amj,
    author = "Favaro, Luigi and Plehn, Tilman and Qu, Huilin and Spinner, Jonas",
    title = "{Virtues and Vices of Equivariant Transformers}",
    eprint = "2608.02735",
    archivePrefix = "arXiv",
    primaryClass = "hep-ph",
    reportNumber = "IRMP-CP3-26-23",
    month = "8",
    year = "2026"
}

@article{Breso-Pla:2026tlz,
    author = "Breso-Pla, Victor and Greif, Kevin and Mikuni, Vinicius and Nachman, Benjamin and Plehn, Tilman and Wamorkar, Tanvi and Whiteson, Daniel",
    title = "{Encoding physics at the precision frontier}",
    eprint = "2603.08802",
    archivePrefix = "arXiv",
    primaryClass = "hep-ph",
    doi = "10.1103/9hs2-prs5",
    journal = "Phys. Rev. D",
    volume = "114",
    number = "3",
    pages = "036034",
    year = "2026"
}

@article{Petitjean:2025zjf,
    author = {Petitjean, Antoine and Plehn, Tilman and Spinner, Jonas and K{\"o}the, Ullrich},
    title = "{Economical Jet Taggers -- Equivariant, Slim, and Quantized}",
    eprint = "2512.17011",
    archivePrefix = "arXiv",
    primaryClass = "hep-ph",
    reportNumber = "IPPP/25/93",
    month = "12",
    year = "2025"
}

@article{Favaro:2025pgz,
    author = "Favaro, Luigi and Gerhartz, Gerrit and Hamprecht, Fred A. and Lippmann, Peter and Pitz, Sebastian and Plehn, Tilman and Qu, Huilin and Spinner, Jonas",
    title = "{Lorentz-Equivariance without Limitations}",
    eprint = "2508.14898",
    archivePrefix = "arXiv",
    primaryClass = "hep-ph",
    doi = "10.21468/SciPostPhysCore.9.3.058",
    journal = "SciPost Phys. Core",
    volume = "9",
    pages = "058",
    year = "2026"
}

@article{Bahl:2025jtk,
    author = "Bahl, Henning and Fuchs, Elina and Menen, Marco and Plehn, Tilman",
    title = "{$\mathcal{CP}$-analyses with symbolic regression}",
    eprint = "2507.05858",
    archivePrefix = "arXiv",
    primaryClass = "hep-ph",
    doi = "10.21468/SciPostPhys.20.2.040",
    journal = "SciPost Phys.",
    volume = "20",
    number = "2",
    pages = "040",
    year = "2026"
}

@article{Butter:2024zbd,
    author = "Butter, Anja and Charton, Fran{\c{c}}ois and Villadamigo, Javier Mari{\~n}o and Ore, Ayodele and Plehn, Tilman and Spinner, Jonas",
    title = "{Extrapolating jet radiation with autoregressive transformers}",
    eprint = "2412.12074",
    archivePrefix = "arXiv",
    primaryClass = "hep-ph",
    doi = "10.21468/SciPostPhys.20.1.004",
    journal = "SciPost Phys.",
    volume = "20",
    number = "1",
    pages = "004",
    year = "2026"
}

@article{Brehmer:2024yqw,
    author = "Brehmer, Johann and Bres{\'o}, V{\'\i}ctor and de Haan, Pim and Plehn, Tilman and Qu, Huilin and Spinner, Jonas and Thaler, Jesse",
    title = "{A Lorentz-equivariant transformer for all of the LHC}",
    eprint = "2411.00446",
    archivePrefix = "arXiv",
    primaryClass = "hep-ph",
    reportNumber = "MIT-CTP/5802",
    doi = "10.21468/SciPostPhys.19.4.108",
    journal = "SciPost Phys.",
    volume = "19",
    number = "4",
    pages = "108",
    year = "2025"
}

@article{Ackerschott:2023nax,
    author = "Ackerschott, Jona and Barman, Rahool Kumar and Gon{\c{c}}alves, Dorival and Heimel, Theo and Plehn, Tilman",
    title = "{Returning CP-observables to the frames they belong}",
    eprint = "2308.00027",
    archivePrefix = "arXiv",
    primaryClass = "hep-ph",
    doi = "10.21468/SciPostPhys.17.1.001",
    journal = "SciPost Phys.",
    volume = "17",
    number = "1",
    pages = "001",
    year = "2024"
}

@article{Butter:2023fov,
    author = "Butter, Anja and Huetsch, Nathan and Palacios Schweitzer, Sofia and Plehn, Tilman and Sorrenson, Peter and Spinner, Jonas",
    title = "{Jet diffusion versus JetGPT {\textendash} Modern networks for the LHC}",
    eprint = "2305.10475",
    archivePrefix = "arXiv",
    primaryClass = "hep-ph",
    doi = "10.21468/SciPostPhysCore.8.1.026",
    journal = "SciPost Phys. Core",
    volume = "8",
    pages = "026",
    year = "2025"
}

@article{Plehn:2022ftl,
    author = "Plehn, Tilman and Butter, Anja and Dillon, Barry and Heimel, Theo and Krause, Claudius and Winterhalder, Ramon",
    title = "{Modern Machine Learning for LHC Physicists}",
    eprint = "2211.01421",
    archivePrefix = "arXiv",
    primaryClass = "hep-ph",
    month = "11",
    year = "2022"
}

@article{Butter:2021rvz,
    author = "Butter, Anja and Plehn, Tilman and Soybelman, Nathalie and Brehmer, Johann",
    title = "{Back to the formula - LHC edition}",
    eprint = "2109.10414",
    archivePrefix = "arXiv",
    primaryClass = "hep-ph",
    doi = "10.21468/SciPostPhys.16.1.037",
    journal = "SciPost Phys.",
    volume = "16",
    number = "1",
    pages = "037",
    year = "2024"
}

@article{Brivio:2019ius,
    author = "Brivio, Ilaria and Bruggisser, Sebastian and Maltoni, Fabio and Moutafis, Rhea and Plehn, Tilman and Vryonidou, Eleni and Westhoff, Susanne and Zhang, C.",
    title = "{O new physics, where art thou? A global search in the top sector}",
    eprint = "1910.03606",
    archivePrefix = "arXiv",
    primaryClass = "hep-ph",
    reportNumber = "P3H-19-036, CERN-TH-2019-193",
    doi = "10.1007/JHEP02(2020)131",
    journal = "JHEP",
    volume = "02",
    pages = "131",
    year = "2020"
}

@article{Brehmer:2017lrt,
    author = "Brehmer, Johann and Kling, Felix and Plehn, Tilman and Tait, Tim M. P.",
    title = "{Better Higgs-CP Tests Through Information Geometry}",
    eprint = "1712.02350",
    archivePrefix = "arXiv",
    primaryClass = "hep-ph",
    reportNumber = "UCI-HEP-TR-2017-14",
    doi = "10.1103/PhysRevD.97.095017",
    journal = "Phys. Rev. D",
    volume = "97",
    number = "9",
    pages = "095017",
    year = "2018"
}

@article{Plehn:2010st,
    author = "Plehn, Tilman and Spannowsky, Michael and Takeuchi, Michihisa and Zerwas, Dirk",
    title = "{Stop Reconstruction with Tagged Tops}",
    eprint = "1006.2833",
    archivePrefix = "arXiv",
    primaryClass = "hep-ph",
    doi = "10.1007/JHEP10(2010)078",
    journal = "JHEP",
    volume = "10",
    pages = "078",
    year = "2010"
}

@article{Plehn:2009rk,
    author = "Plehn, Tilman and Salam, Gavin P. and Spannowsky, Michael",
    title = "{Fat Jets for a Light Higgs}",
    eprint = "0910.5472",
    archivePrefix = "arXiv",
    primaryClass = "hep-ph",
    reportNumber = "KA-TP-12-2009",
    doi = "10.1103/PhysRevLett.104.111801",
    journal = "Phys. Rev. Lett.",
    volume = "104",
    pages = "111801",
    year = "2010"
}

@article{Plehn:2001nj,
    author = "Plehn, Tilman and Rainwater, David L. and Zeppenfeld, Dieter",
    title = "{Determining the Structure of Higgs Couplings at the LHC}",
    eprint = "hep-ph/0105325",
    archivePrefix = "arXiv",
    reportNumber = "MADPH-01-1229, FERMILAB-PUB-01-067-T",
    doi = "10.1103/PhysRevLett.88.051801",
    journal = "Phys. Rev. Lett.",
    volume = "88",
    pages = "051801",
    year = "2002"
}
\end{document}